\documentclass[fleqn,10pt]{wlscirep}
\usepackage[utf8]{inputenc}
\usepackage[T1]{fontenc}

\title{OxyJet: Design and Evaluation of A Low-Cost Precision Venturi Based Continuous Positive Airway Pressure (CPAP) System}

\author{Md. Kawsar Ahmed, Meemnur Rashid, Kaisar Ahmed Alman, Farhan Muhib, Saeedur Rahman, and Taufiq Hasan}
\affil{Department of Biomedical Engineering, Bangladesh University of Engineering \& Technology, Dhaka, Bangladesh}

\begin{abstract}
The Covid-19 pandemic has strained the hospital systems in many countries in the world, especially in the developing countries. In many low-resource hospitals, severely ill hypoxemic Covid-19 patients are treated with various forms of low-flow oxygen therapy (0-15 L/min), including interfaces such as, nasal cannula, hudson mask, venturi-mask, and non-rebreather masks. When the oxygen saturation of patients cannot be maintained on 15L/min of pure oxygen flow, international treatment guidelines suggest non-invasive positive pressure ventilation (NIPPV) or high-flow nasal oxygenation (HFNO) as the next stage of treatment to avoid invasive mechanical ventilation (IMV). However, administering high-flow nasal oxygenation (HFNO) in the general wards of a low-resource hospital is difficult due to a number of factors, including difficulty in operation, unavailability of electric power outlets and frequent maintenance. Therefore, in many cases the highest level of care a patient receives in the general ward is 15L/min of oxygen on a Non-Rebreather Mask. With shortage of Intensive Care Unit (ICU) beds, this is a major problem since intermediate forms of treatments are simply not available at an affordable cost.
To address this gap, we have developed the \emph{OxyJet CPAP}, a low-cost CPAP system specifically designed for low and middle-income country (LMIC) hospitals. The device is a precision venturi-based flow-generator capable of providing up to 60L/min of flow. The device utilizes the mechanics of a jet pump driven by high-pressure oxygen to increase the volumetric flow rate by entraining atmospheric air. Using a dual-flowmeter, the fraction of inspired oxygen ($FiO_2$) can be attained between 40 - 100\%. Consisting of a traditional 22mm breathing circuit, a non-vented CPAP mask, and a Positive End-Expiratory Pressure (PEEP) valve, the CPAP can provide positive pressures between 5-20 cm $H_2O$. The device is manufactured using local 3D printing and workshop facilities. The pressure and flow characteristics of OxyJet are equivalent to existing CPAP devices and the technical evaluations meet the requirement of the UK-MHRA Rapidly Manufacturable CPAP systems (RMCPAPS) guideline.
\end{abstract}
\begin{document}

\flushbottom
\maketitle
%
%
\thispagestyle{empty}

\section{INTRODUCTION}
The coronavirus disease 2019 (COVID-19) caused by severe acute respiratory syndrome coronavirus 2 (SARS-CoV-2) has been devastating to the communities worldwide. In Bangladesh, the first confirmed case of the virus was found in March, 2020. According to the World Health Organization \cite{world2020clinical}, most COVID-19 patients have uncomplicated or mild illness (81\%) while some develop severe illness requiring hospitalization and oxygen therapy (14\%). Finally, about 5\% of patients become critically ill requiring ICU admission and possibly mechanical ventilation. COVID-19 primarily affects the respiratory system while severe cases may lead to acute respiratory failure \cite{world2020clinical}, requiring oxygen therapy. According WHO guidelines, supplemental oxygen therapy should be administered to any patient showing emergency signs or having SpO2 $<$ 90\%. Oxygen delivery devices can be progressively used in the following order:  nasal cannula (upto 5 L/min), Venturi mask (6–10 L/min), and non-rebreather masks (10–15 L/min) with a target of achieving an SpO2 $>$ 90\% (non-pregnant adults) \cite{world2020clinical,whittle2020respiratory}. If the patient’s condition worsens, WHO recommends a trial of High-Flow Nasal Oxygenation (HFNO) or non-invasive positive pressure ventilation (NIPPV) including: continuous positive airway pressure (CPAP) and bilevel positive airway pressure (BiPAP) \cite{rhodes2017surviving,rimensberger2015ventilatory}. 
The patient is recommended for intubated mechanical ventilation as the last resort only if these non-invasive treatment pathways have failed \cite{covid19guidelineBD}.

With a population of 160 million, the number of severe COVID-19 patients requiring hospitalization can be very high in Bangladesh if the pandemic becomes widespread. Moreover, there is a severe shortage of key medical equipment to address the situation, especially in the underserved regions of Bangladesh. Firstly, there is a lack of infrastructure for oxygen supply in different regions of the country and most government hospitals do not have centralized oxygen supply \cite{oxygen2020news}. Secondly, there is only 0.72 ICU beds per 100,000 people in Bangladesh \cite{abdullah2020number}, compared to 34.7, 12.5, and 2.3 for the United States, Italy, and India, respectively \cite{mccarthy2020countries}. In addition, with a lack of skilled ICU doctors and nurses, it is very difficult to treat severely ill patients even if adequate medical equipment is made available. The mortality rate for patients that have been placed under a mechanical ventilator was found to be very high \cite{namendys2020respiratory}, and accordingly the Bangladesh COVID-19 guidelines recommend trying to avoid intubation at all costs \cite{covid19guidelineBD}.

In recent times, HFNO devices have been found to be effective in treating hypoxemic COVID-19 patients while reducing the need for mechanical ventilation \cite{rochwerg2017official,slessarev2020patient}. Although the use of CPAP and BiPAPs are debated for COVID-19 \cite{namendys2020respiratory}, studies have also shown the effective use of CPAPs to reduce ICU demand and mortality \cite{lawton2020reduced,ashish2020early}. One study concluded that, using HFNO vs NIPPV for critical COVID-19 patients did not yeild significantly differnt outcomes in terms of intubation rate and mortality \cite{duan2020use}. While WHO guideline \cite{world2020clinical} does not specifically mention an exact sequence of intermediate therapies (HFNO and NIPPV) to administer before intubation, the United Kingdom National Health Service (NHS) guidelines particularly mention the preference of NIPPV over HFNO \cite{nhs1559guidance}. According to the NHS guidelines, non-invasive ventilation (NIV) may be used as (i) a ceiling of treatment, (ii) trial to avoid intubation, or (iii) to facilitate extubation. Non-invasive ventilation (e.g., a CPAP) effectively unloads the respiratory muscle, increases the tidal volume, decreases diaphragmatic work of breathing, and thus reduces hypoxia by maintaining a PEEP. A device capable of providing a high enough flow while maintaining the PEEP ensures that the patient receives a sufficient amount of oxygen, prevents the alveoli from collapsing while recruiting new alveoli for respiration \cite{And}.  However, the risk of aerosolization is well known for CPAP and BiPAP therapies when using traditional vented masks \cite{simonds2010evaluation}. This risk can be minimized by using oxygen hood/helmet-based delivery  or by using a negative-pressure room \cite{ing2020role,marini2020management,radovanovic2020helmet}. In \cite{gattinoni2020covid}, it was hypothesized that two distinct phenotypes, L and H exists for the COVID-19 respiratory failure.  The ``Type H" respiratory failure, characterized by more extensive consolidation and high lung elastance, is more responsive to a high Positive End Expiratory Pressure (PEEP) \cite{marini2020management}. 

In the context of LMICs, various limitations and resource contraints must be taken into account while deciding the most effective clinical treatment pathway. Although HFNO devices are effective, they are expensive (above \$5,000 for reputable brands) and difficult to import. Due to a lack of regulation, local suppliers may ask for more than double the original price of the machines. In this situation, the use of low-cost CPAP devices as an intermediary treatment maybe considered in order to avoid intubation \cite{nhs1559guidance,covid19guidelineBD}. Since CPAP is an aerosol generating procedure \cite{whittle2020respiratory}, either negative-pressure room or non-vented CPAP masks or oxygen hoods / helmets need to be used. Since the non-vented CPAP masks are low cost and more readily available for purchase in the market, this is a viable option in the context of Bangladesh. Hence, we propose to develop a low-cost CPAP system \emph{"Oxyjet CPAP"} suitable for use in LMIC settings. 

\begin{figure}[t]
    \centering
    \includegraphics[width=\linewidth]{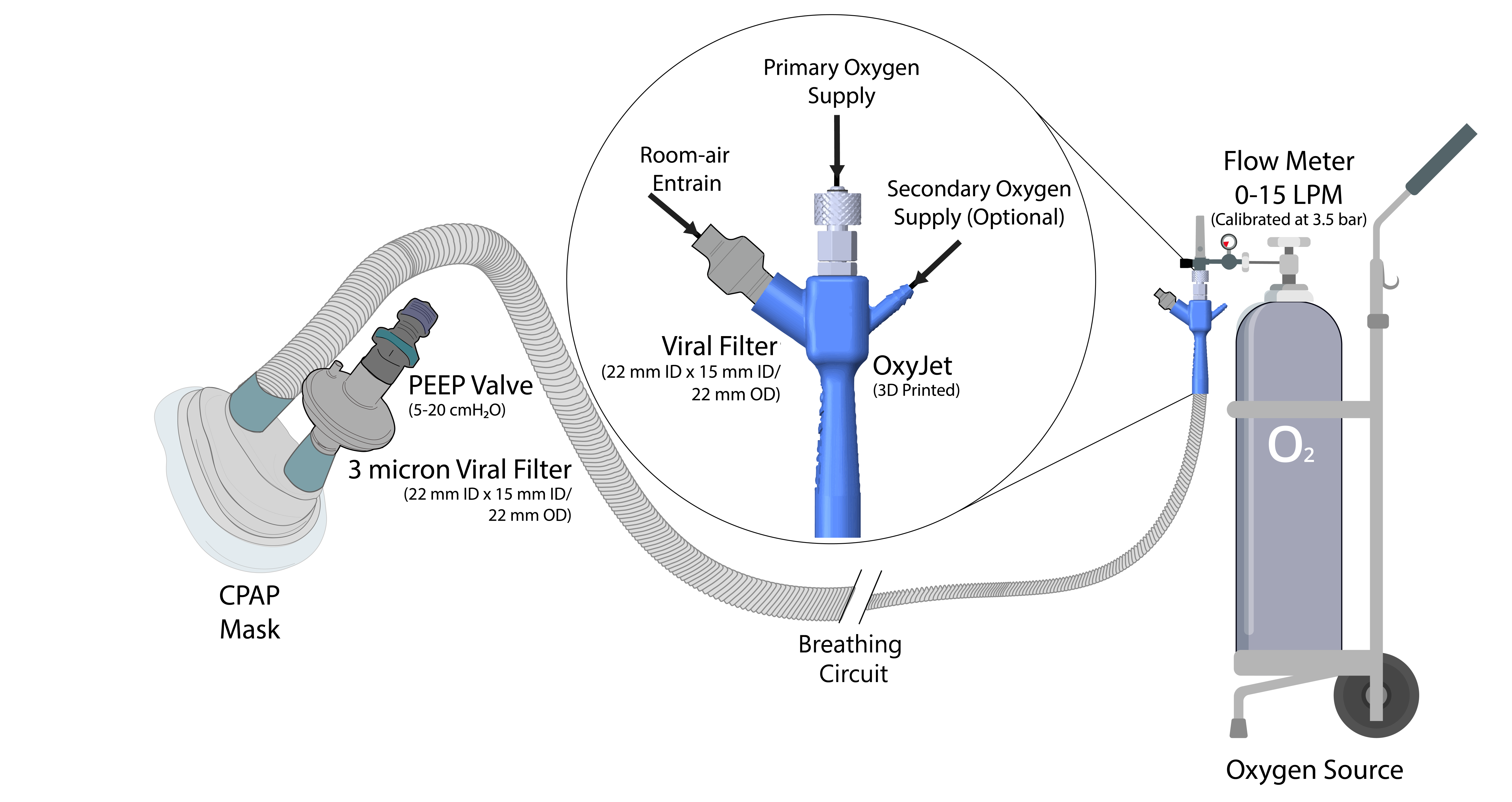}
    \caption{OxyJet CPAP System Overview. 
    The Oxyjet flow generator readily attaches with a standard flowmeter (15L/min). The air inlet of the flow generator is protected with a viral filter. The secondary oxygen inlet of the flow generator enables supply of additional oxygen from a secondary flowmeter (50L/min) to increase the oxygen concentration up to 100\%. The outlet of the flow generator is attached to a long breathing circuit at one end and the inhalation port of patient interface (face-mask or helmet) at the other end. The exhalation port of the patient interface is connected to a viral filter followed by a PEEP valve.}\label{fig:system_overview}    
\end{figure}

\section{OXYJET CPAP System}
\label{sec:cpap}

OxyJet CPAP is a low cost CPAP system developed with an aim of early intervention and preventing patients from going to the ICU. Almost all of its components are off the shelf, except the OxyJet flow generator. 
The OxyJet flow generator is a pressure driven flow generator which functions on the basis of jet mixing principle and deliver a continuous flow of oxygenated air to the patient and maintains a positive pressure inside the patient's mask. The overall system has the following features:
\begin{itemize}
    \item Simple and easy to use
    \item Works without electric power 
    \item Compatible with standard oxygen flowmeter ports
    \item Non-vented CPAP mask/hood with viral filter minimizes aerosol generation and contamination in the hospital
    \item Provide high flow of up to 61 L/min
    \item Provide positive end expiratory pressure (PEEP) between 5 - 20 cm $ H_2O $
    \item Dual oxygen flowmeters can provide a fraction of inspired oxygen (\%FiO$_2$) of up to 100\%
\end{itemize}

\subsection{Device Operation}
\begin{figure}[ht]
    \centering
    \includegraphics[width = \linewidth]{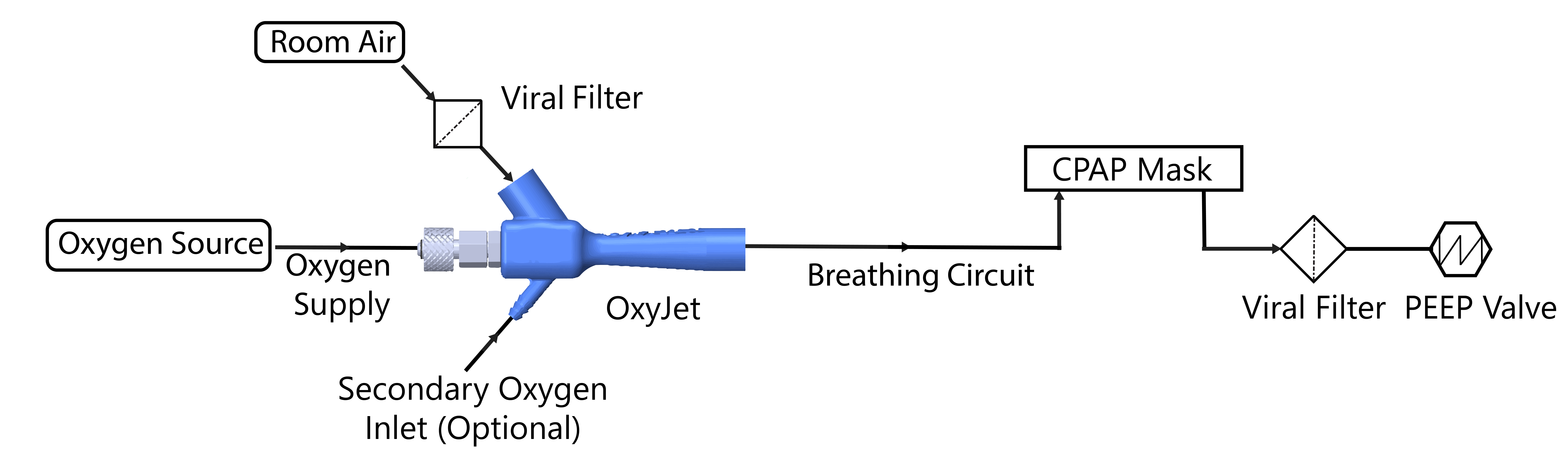}
    \caption{A schematic diagram of the OxyJet CPAP breathing circuit.}
    \label{fig:flowchart}
\end{figure}

The oxyjet flow generator is attached to an oxygen source securely with a DISS connector. The flow generator consists of three inlets: (i) primary oxygen inlet, (ii) air inlet, and (iii) secondary oxygen inlet as shown in Fig \ref{fig:system_overview}. When pressurized oxygen is supplied through the primary inlet it gains high-velocity as it passes through a fine nozzle. This driving flow entrains room air through the air inlet. Inside the device, the room air is mixed with the high-velocity oxygen and an oxygenated high-flow of air is channeled through the breathing circuit. The viral filter at the air inlet filters out viruses and dust particles of the entrained room air. A tightly fitted face mask (non-vented) or helmet minimizes aerosolization. The viral filter at the exhalation port decontaminates expiratory air. The adjustable PEEP valve maintains a fixed positive end expiratory pressure (5-20 cm $H_2O$) as required by patients. A flowchart of the CPAP system is shown in Fig. \ref{fig:flowchart}.

\subsection{OxyJet Flow Generator}
\label{sec:1}
\begin{figure} [ht]
        \centering
        \begin{tabular}{c c}
        \includegraphics[width=0.2\textwidth]{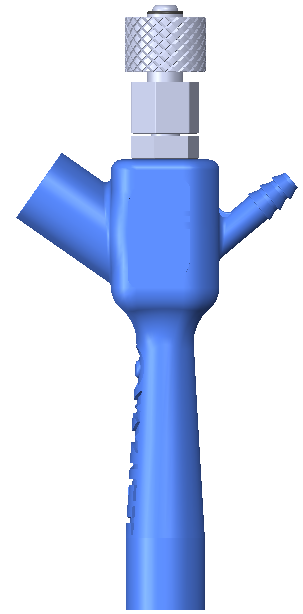}     &  \includegraphics[width=0.78\textwidth]{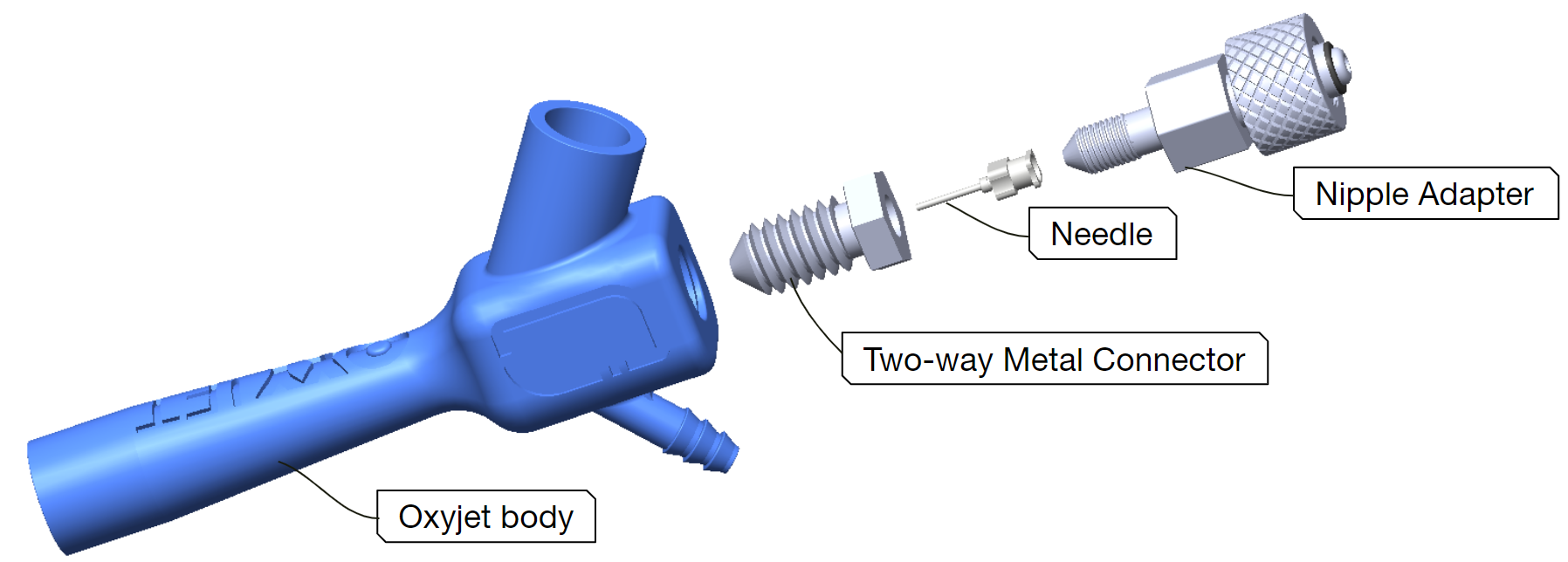}\\
         (a)    & (b)
        \end{tabular}
        \caption{Illustrations of (a) the oxyjet flow generator and (b) its exploded view.}\label{oxyjet_flow_generator}
\end{figure}

Oxyjet flow generator is a specially designed gas ejector (Fig \ref{oxyjet_flow_generator}), produced from locally available materials, to minimize manufacturing complications without hampering the device performance. The flow generator comprises of four main components, a DISS nipple adapter with a metal knob, a primary nozzle (needle), a specially designed two-way metal connector, and the oxyjet body.

\subsubsection{Primary Nozzle (needle)}
It is the heart of the flow generator. It utilizes high pressure of the O$_2$ source to accelerates the driving flow to a sonic/supersonic condition. There are several important considerations while designing and developing the primary nozzle. The nozzle has to withstand high pressure (3.5 bar or 50 psi) of the O$_2$ source. The nozzle has to be reproducible with high precision, as a slight deviation can abruptly change the device performance. Material of the nozzle is also important, since the nozzle has to come in direct contact with the O$_2$ and air inhaled by patients.

\begin{figure} [ht]
        \centering
        \begin{tabular}{c c}
        \includegraphics[width=0.25\textwidth]{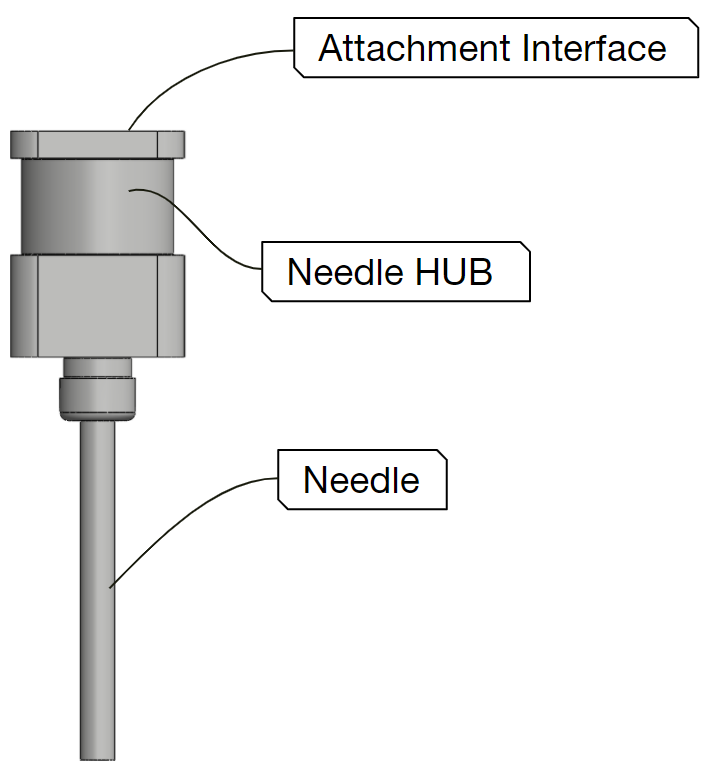}     &  \includegraphics[width=0.3\textwidth]{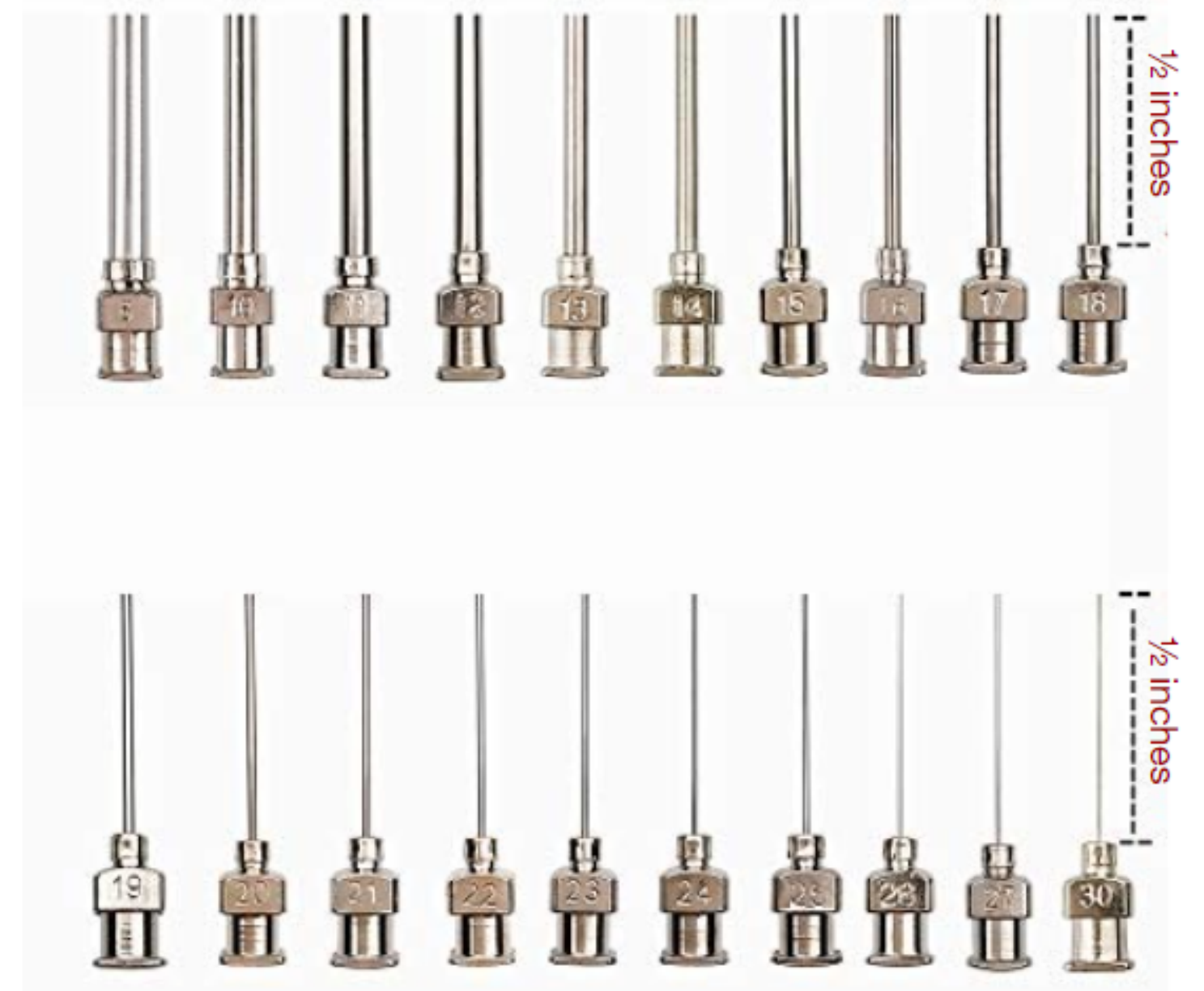}\\
         (a)    & (b)
        \end{tabular}
        \caption{Illustration of (a) an industrial grade blunt tip stainless steel dispensing needle and (b) it's varying gauges}\label{ss_needle}
\end{figure}

Taking all these into consideration, design and development of the primary nozzle would be a challenging task in a low resource facility. Thus, we came up with the idea of repurposing an industrial grade blunt tip stainless steel dispensing needle as the primary nozzle (Fig \ref{ss_needle}). The needles can withstand high pressure upto 100 psi (from specification) \cite{needle}, which is twice the maximum pressure supplied by the O$_2$ source. The needles are off the shelf and are manufactured with high precision, with little to no deviation in their inner diameter \cite{needlegauge}, this solves the reproducibility issue . Moreover, the needles are made of stainless steel or nickel plated brass \cite{needle}, which are ideal materials of the oxygen therapy delivery system\cite{fisher}. Thus, the needles fulfill all of the important considerations.

Finally, we have to test the performance of these needles with different gauges and the flow generator itself to evaluate the feasibility of our concept and determine the optimum gauge. We have performed one-dimensional flow analysis and computational fluid dynamics (CFD) simulations to determine needle gauges that suit our device requirements.

For one-dimensional isentropic flow, the local pressure, temperature and density is related to their corresponding values at stagnation condition by the following relations.

\begin{equation}
    \frac{p_0}{p} = \left(1+\frac{\gamma - 1}{2} M^2\right)^\frac{\gamma}{\gamma-1} \tag{2.2.1.1}
 \end{equation}  
\begin{equation}
    \frac{T_0}{T} = \left( 1+\frac{\gamma - 1}{2} M^2 \right) \tag{2.2.1.2}
\end{equation}  

\begin{equation}
    \frac{\rho_0}{\rho} = \left(1+\frac{\gamma - 1}{2} M^2\right)^\frac{1}{\gamma-1} \tag{2.2.1.3}
\end{equation}

Now, using the relation (2.1.1) we have compute the minimum stagnation pressure required across any needle gauge for choked flow (M=1), which is $p_0 $ = 191801.05 Pa or 1.918 bar (for the value of $\gamma = 1.4$ and local pressure $p $ = 1 atm or 101325 Pa). Hence, any stagnation pressure greater than 1.918 bar (absolute) across a nozzle should be enough to produce choked flow. 

In our device setup, we have used the existing standard pressure regulator and flowmeter arrangements that come with the O$_2$ source. The maximum operating point of these standard flowmeters are 15 L/min (measurable) flowrate of pure oxygen and 3.5 bar or 50 psi stagnation pressure.

Using the equation (2.1.4), we have performed some analysis to determine the choked flow rate through nozzles for one-dimensional isentropic condition with the maximum operating point of the O$_2$ source at inlet.

\begin{equation}
\dot{m} = \frac{p_0 A^*}{\sqrt{T_0}} \sqrt{\frac{\gamma}{R}\left(\frac{2}{\gamma+1}\right)^\frac{\gamma+1}{\gamma-1}} \tag{2.2.1.4}
\end{equation}
Using the choked flow equation (2.2.1.4), a relation (2.2.1.5) can be derived between flow rate (at SLPM) and needle gauge (G).

\[\ \dot{m} = \rho_{slpm} Q_{slpm} =\rho^* u^* A^* = \frac{p_0 A^*}{\sqrt{T_0}} \sqrt{\frac{\gamma}{R}\left(\frac{2}{\gamma+1}\right)^\frac{\gamma+1}{\gamma-1}} \]

\begin{equation}
Q_{slpm} =\frac{p_0 \left(\frac{\pi d^2}{4}\right)}{\rho_{slpm} \sqrt{T_0}} \sqrt{\frac{\gamma}{R}\left(\frac{2}{\gamma+1}\right)^\frac{\gamma+1}{\gamma-1}} \tag{2.2.1.5}
\end{equation}

Where $ A^* = A_t = \frac{\pi d^2}{4} $ at critical condition and "d" is the inner diameter of a needle labeled with gauge (G).

\begin{figure}[!h]
    \centering
    \includegraphics[width=0.6\textwidth]{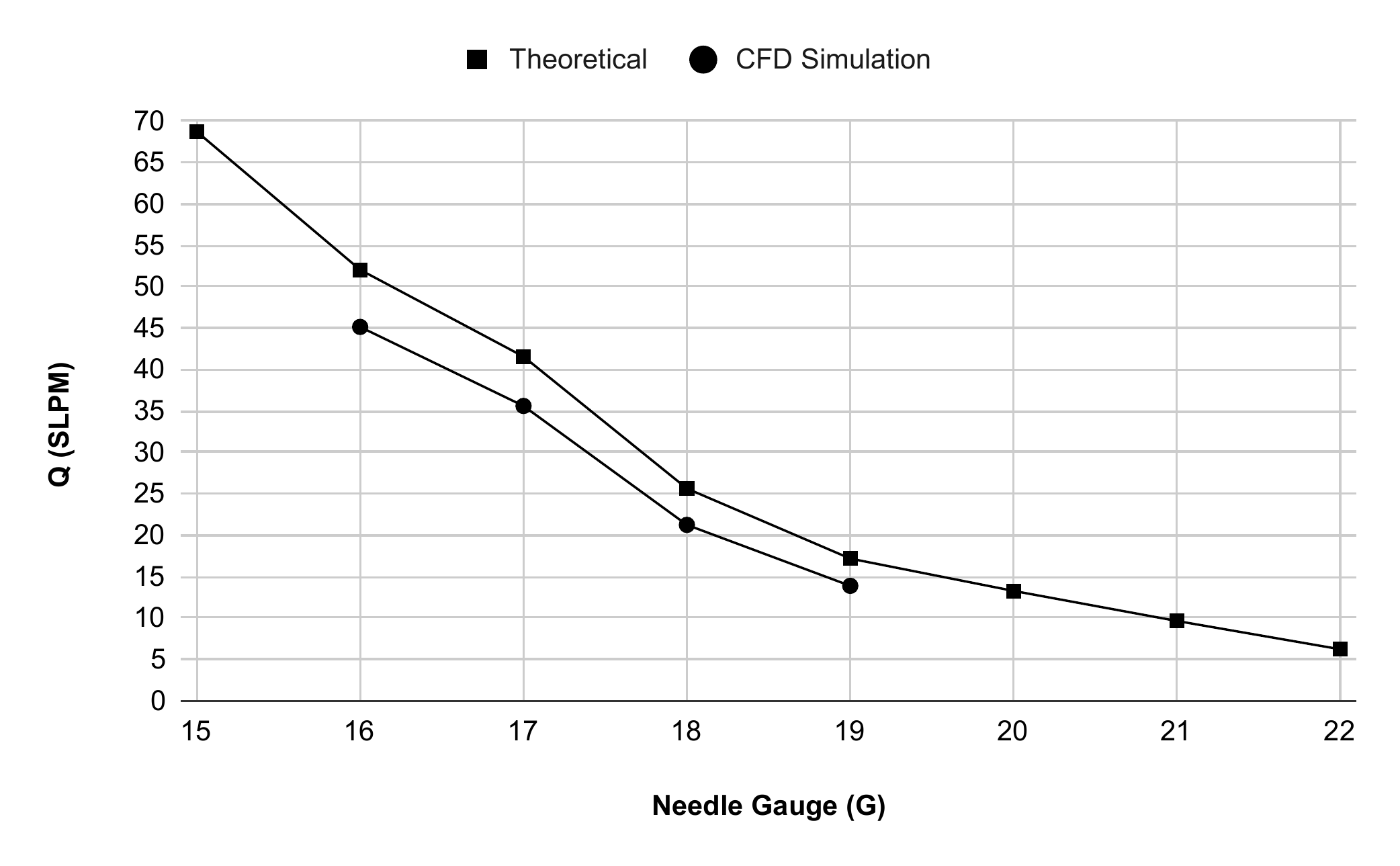}
    \caption{Maximum allowable flow (SLPM) vs Needle gauge for $ p_0 $ = 3.5 bar and $ T_0 $ = 300K.The CFD simulation results are in close agreement with one-dimensional flow analysis. However, the CFD simulation curve is always below the one-dimensional flow curve, which is acceptable because CFD simulation takes many factors (like viscosity, wall roughness etc) into consideration.}
    \label{q_vs_needle}
\end{figure}

We have also performed some Computational Fluid Dynamics (CFD) analysis and compared with our theoretical findings. Fig \ref{q_vs_needle}, illustrates the comparison of the maximum allowable flow (choked flow) vs needle gauge derived using equation (2.1.5) and CFD analysis, with the boundary conditions: stagnation pressure $p_0$ = 3.5 bar or 50 psi (gauge) and stagnation temperature $T_0$ = 300K. In contrast to the one-dimensional isentropic theoretical analysis, three dimensional turbulence, viscous flow with wall roughness was assumed for the CFD simulations. SST-kw was used as the turbulence model.

\begin{figure} [!ht]
        \centering
        \begin{tabular}{c c}
        \includegraphics[width=0.49\textwidth]{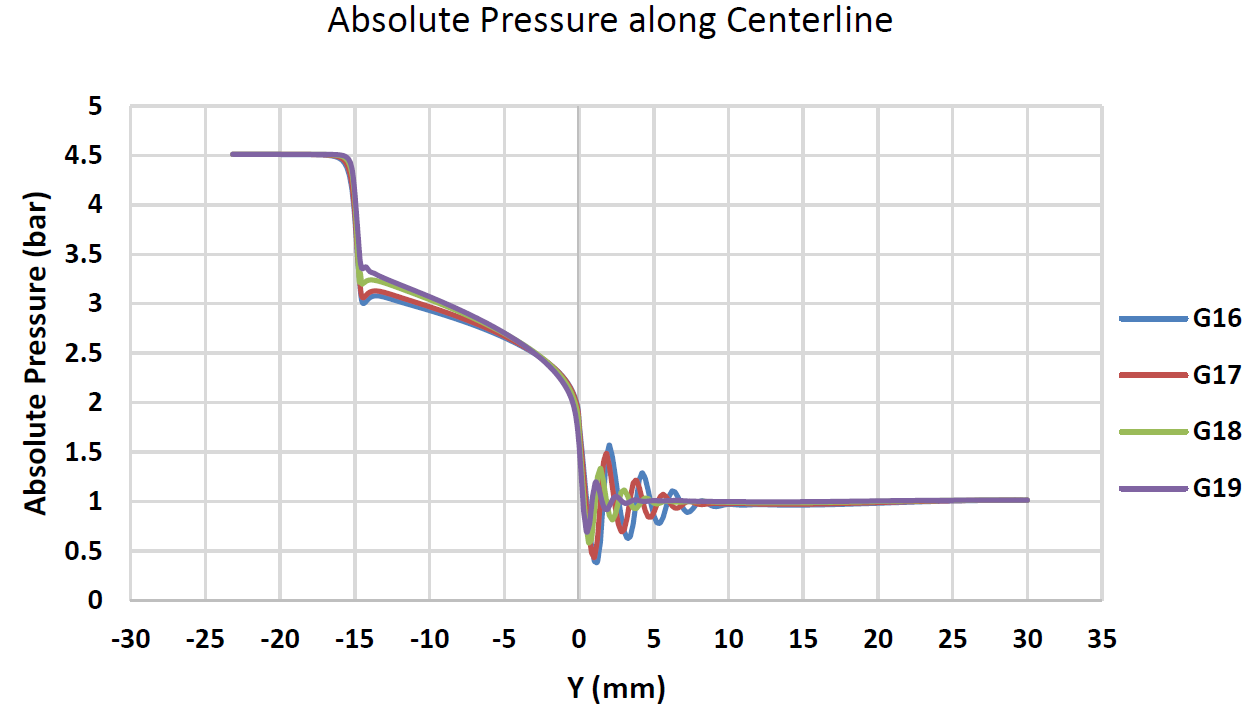}     &  \includegraphics[width=0.49\textwidth]{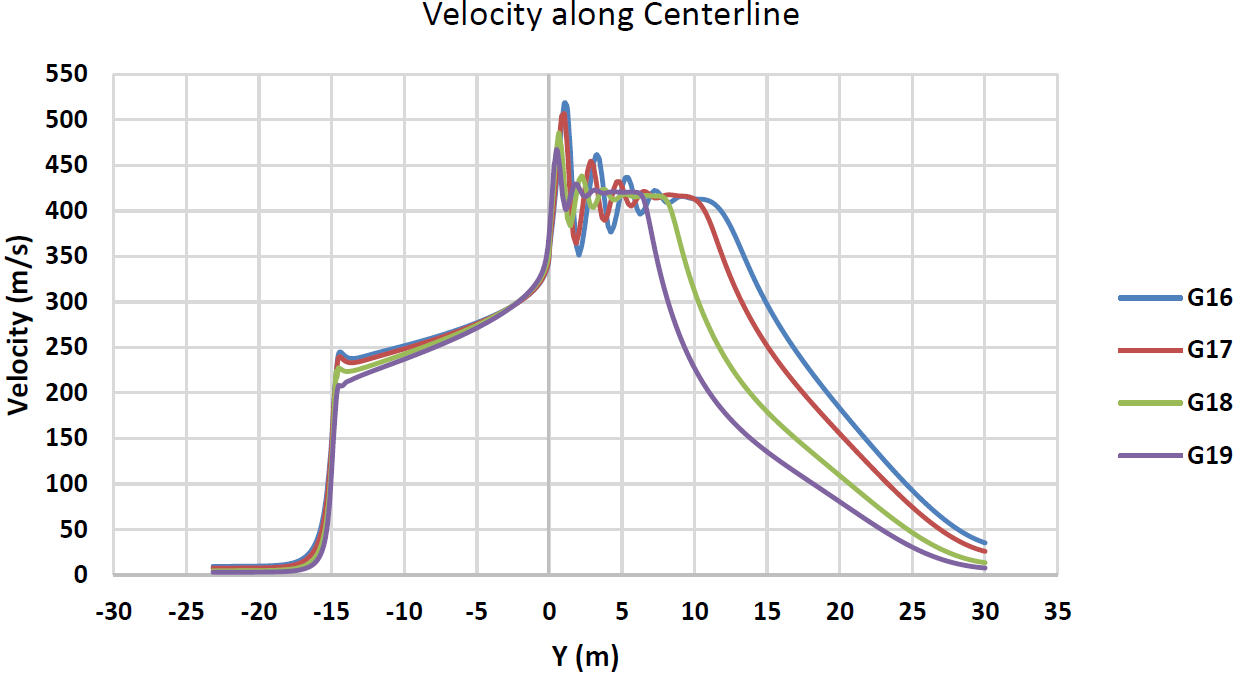}\\
         (a)    &  (b)
        \end{tabular}
        \caption{Comparison of the variation of (a) center-line absolute pressure and (b) center-line velocity with distance for needles with different gauges. Y=0 mm indicates needle exit. CFD simulations include ambient environment at the needle exit. All the needle gauges produce similar curves. 
        The supersonic under-expanded nozzle flow causes sharp pressure drop at the needle exit and produces shock diamond. The amplitude of the pressure oscillation at the needle exit increases as the needle gauge decreases.The center-line velocity curves indicate that decrease in pressure is accompanied by increase in velocity. Velocity oscillations are also observed in the shock diamond and there is a subsequent decrease in velocity as the jet diffuses. However, the amplitude of the velocity oscillation and the length of the region before the start of jet diffusion increases as the needle gauge decreases. Needles of lower gauges produce jets having larger diameter.}\label{c_pressure_velocity}
\end{figure}

Fig \ref{c_pressure_velocity} illustrates the CFD simulation result of the variation of center-line absolute pressure and center-line velocity with distance for needles with different gauges (G16, G17, G18 and G19).

\begin{figure} [ht]
        \centering
        \begin{tabular}{c c}
        \includegraphics[width=0.49\textwidth]{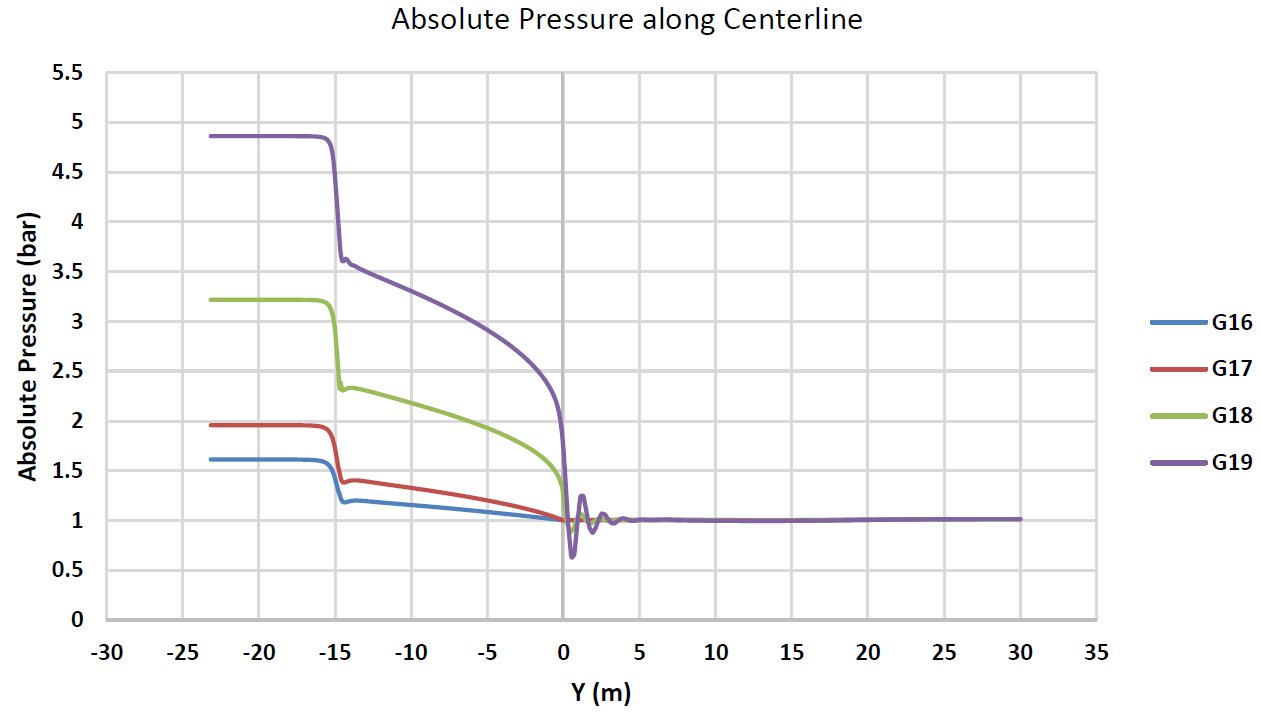}     &  \includegraphics[width=0.49\textwidth]{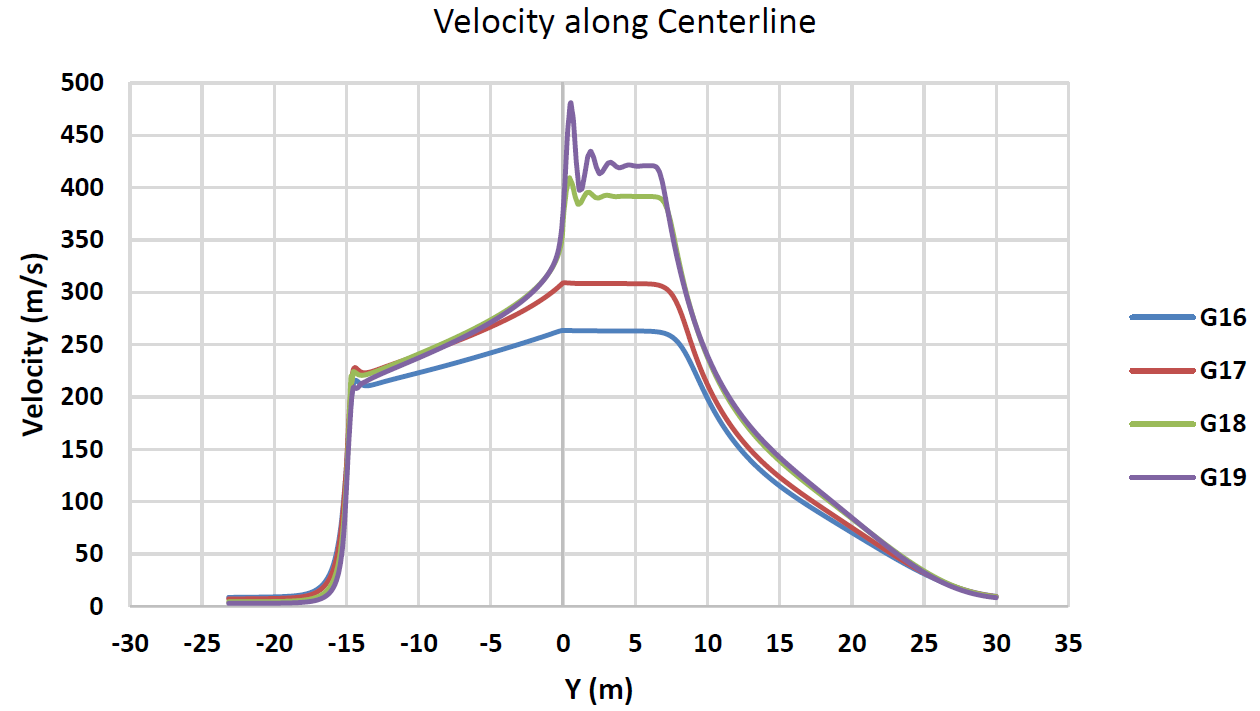}\\
         (a)    & (b)
        \end{tabular}
        \caption{Comparison of the variation of (a) center-line absolute pressure and (b) center-line velocity with distance for needles with different gauges.  For a fixed flowrate (15 SLPM), the center-line absolute pressure and center-line velocity variation curves are different for needles with different gauges. }\label{c_pressure_velocity_p}
\end{figure}

In the next step, we have evaluated the pressure drop across needles with different gauges, for a fixed inlet flow of 15 L/min and inlet stagnation temperature $T_0 $ = 300K using one-dimensional flow analysis and CFD simulation. Fig \ref{c_pressure_velocity_p} illustrates the CFD simulation result of the variation of center-line absolute pressure and center-line velocity with distance for needles with different gauges (G16, G17, G18 and G19).

The needles with gauge 17 and 16 produces subsonic compressible flow (Mach number $<$ 1) whereas the needles with gauge 19 and 18 produces supersonic flow. As the needle gauge increases, the inlet stagnation pressure increases. The gauge pressures are 3.85 bar, 2.2 bar, 0.94 bar and 0.59 bar for needles with gauge 19, 18, 17 and 16 respectively. Moreover, pressure drops sharply at the needle exit with oscillation (shock diamond) for needles with gauge 19 and 18. However, no sharp fall of pressure or pressure oscillation (shock diamond) is observed in case of needles with gauge 17 and 16. The needle exit velocity also increases with needle gauge. Moreover, there is a sharp increase in velocity at the nozzle exit and presence of oscillation (shock diamond) for needles with gauge 19 and 18. However, the needle exit velocity remains subsonic compressible for needles with gauge 17 and 16. Hence, the required inlet stagnation pressure for needles with gauge equal or higher than 18 can be determined using the one-dimensional isentropic flow equation (2.1.5) directly. However, for the needle with gauge lower than 18, which produces subsonic compressible flow we have to deduce an equation from the continuity equation:

\begin{equation} \label{eq1} \tag{2.2.1.6}
\begin{split}
    \dot{m} & = \rho A u = \rho A M c \\
    & = \frac{P}{RT} A M c = \frac{P}{RT} A M \sqrt{\gamma R T} \\
\end{split}
\end{equation}

Using equation (2.2.1.2) and (2.2.1.6) we can deduce
\[\ M^4 + \frac{2}{\gamma - 1} M^2 - \frac{2\dot{m}^2}{P^2A^2}\frac{RT_0}{\gamma(\gamma-1)} = 0 \tag{2.2.1.7} \]
Solving the equation (2.1.7) for needles with gauge lower than 18 with the boundary conditions: stagnation temperature $T_0 $ = 300K and flowrate = 15 SLPM we can determine the real value of mach number M. Now, using the equation (2.1.1) we can determine the stagnation pressure $p_0$. 

\begin{figure}[!h]
    \centering
    \includegraphics[width=0.6\textwidth]{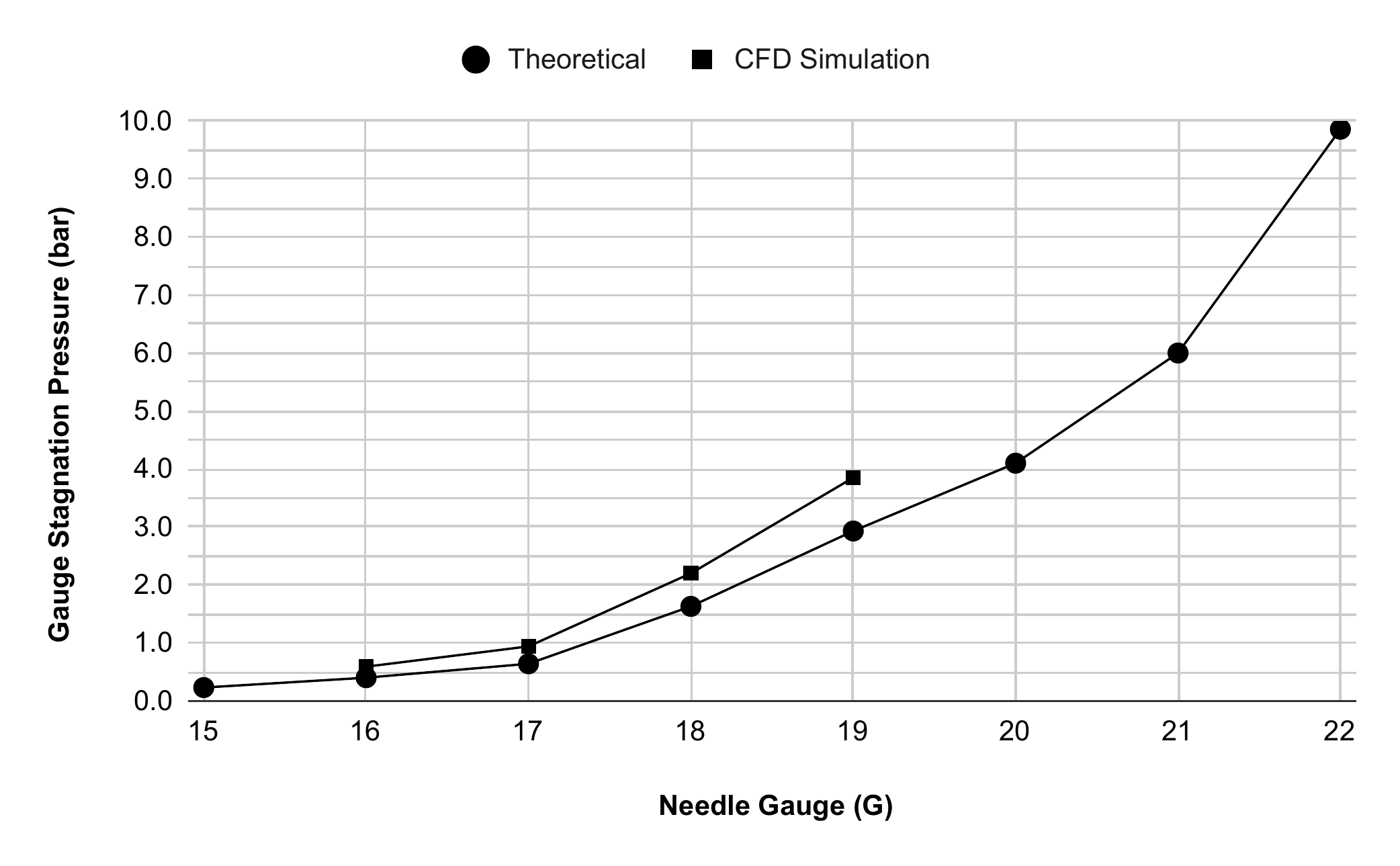}
    \caption{Comparison of the pressure drop across needle of different needle gauges derived from CFD simulation and one-dimensional flow analysis with boundary conditions: Flowrate = 15 LPM and $ T_0 $ = 300K. The CFD simulation results are in close agreement with the theoretical results. However, the pressure drop curve of the CFD simulation is always above the one-dimensional flow analysis curve (theoretical) , because CFD simulation takes many factors (like viscosity, wall roughness etc) into considerations.}
    \label{pr_vs_needle}
\end{figure}

Fig \ref{pr_vs_needle} illustrates the comparison of the required stagnation pressure (gauge) at the needle inlet with different gauges derived from CFD simulation and one-dimensional flow analysis. 
From graph (Fig \ref{q_vs_needle}) it is observed that the needle with gauge 19 is the closes gauge that operates approximately at the maximum operating point (15 L/min flow) utilizing maximum pressure of the O$_2$ source.  Graph (Fig \ref{pr_vs_needle}) also supports the above statement.
However, as we gradually decrease the needle gauges from G19 to G18, G17, or G16, the stagnation pressure at needle inlet $ p_0 $ also decreases for a fixed flowrate (15 LPM) of O$_2$. (from fig \ref{pr_vs_needle}). This suggests that the lower the gauge of the needle, the less it utilizes the pressure delivered by the O$_2$ source.

Based on the above analysis we came to several conclusions: i) Needle with gauge 19 utilizes maximum pressure of the O$_2$ source ii) Needle with gauge lower than 19 utilizes less pressure than the pressure delivered by the O$_2$ source and iii) Needle with gauge G18 and higher attains a supersonic state whereas needle with gauge lower than G18 remains in the subsonic region. 
These analysis helped us understand some basic properties of these needles and how they vary with gauges (G19, G18, G17, and G16).

\subsubsection{DISS Nipple adapter}
\begin{figure}[!h]
    \centering
    \includegraphics[width=0.5\textwidth]{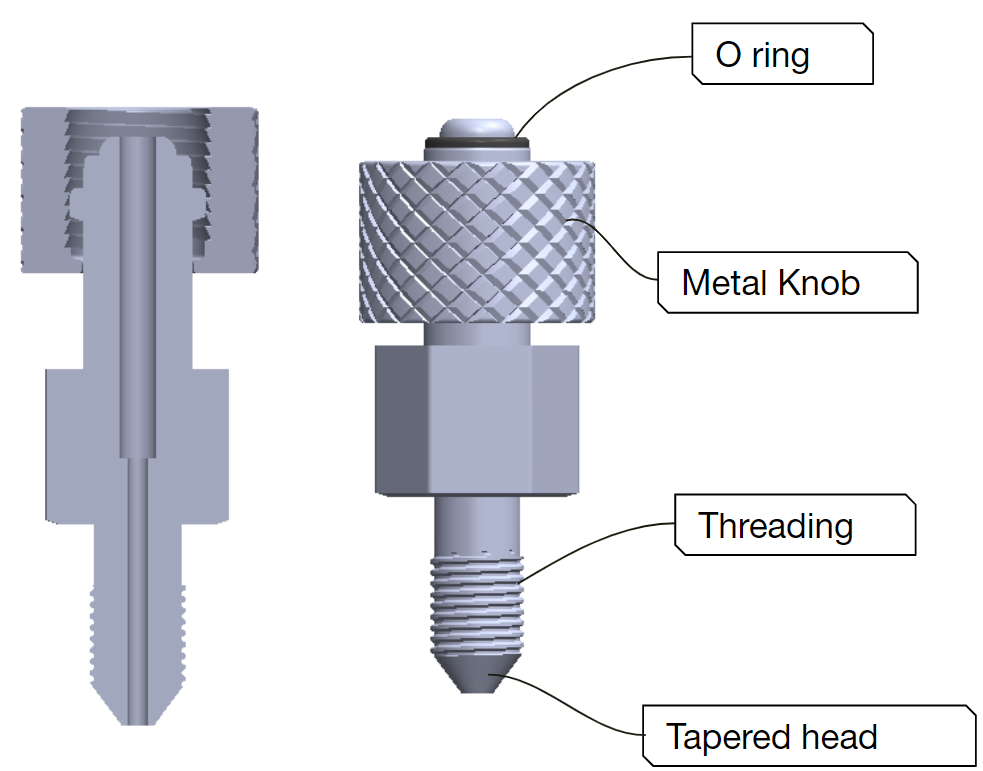}
    \caption{Illustration of an off the shelf DISS Nipple adapter and its cross section}
    \label{nipple_adapter}
\end{figure}

DISS Nipple adapter is an off the shelf component. However, there are several types of oxygen connector and we need a specific type as illustrated in Fig \ref{nipple_adapter}. This adapter is used to connect our flow generator directly to any standard flowmeter. The metal knob at the top is used to attach the adapter with a standard flowmeter and the O-ring ensures no leak. The bottom part of the adapter has threading on the outside wall and a tapered tip. The tapered tip fits perfectly on the needle hub.

\subsubsection{Two-way Metal Connector}
A custom-made two-way metal connector that has a chamber and inner threading that attaches the needle air-tightly with the DISS nipple adapter. There is also outside threading that connects the whole arrangement with the Oxyjet body. 

\begin{figure}[!h]
    \centering
    \resizebox{0.6\textwidth}{!}{\includegraphics{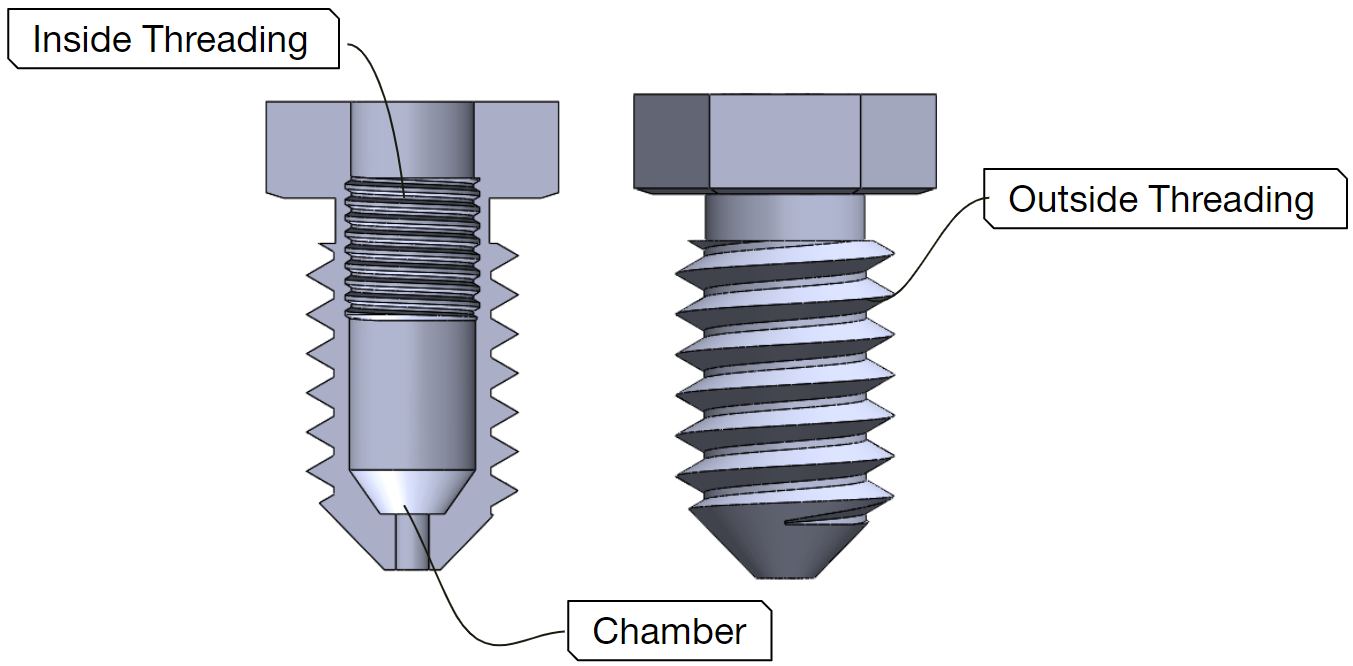}}
    \caption{Illustration of a custom made two-way metal connector and its cross section}
    \label{two-way-connector}
\end{figure}

\subsubsection{Assembled adapters with primary nozzle}
\begin{figure}[!h]
    \centering
    \resizebox{0.4\textwidth}{!}{\includegraphics{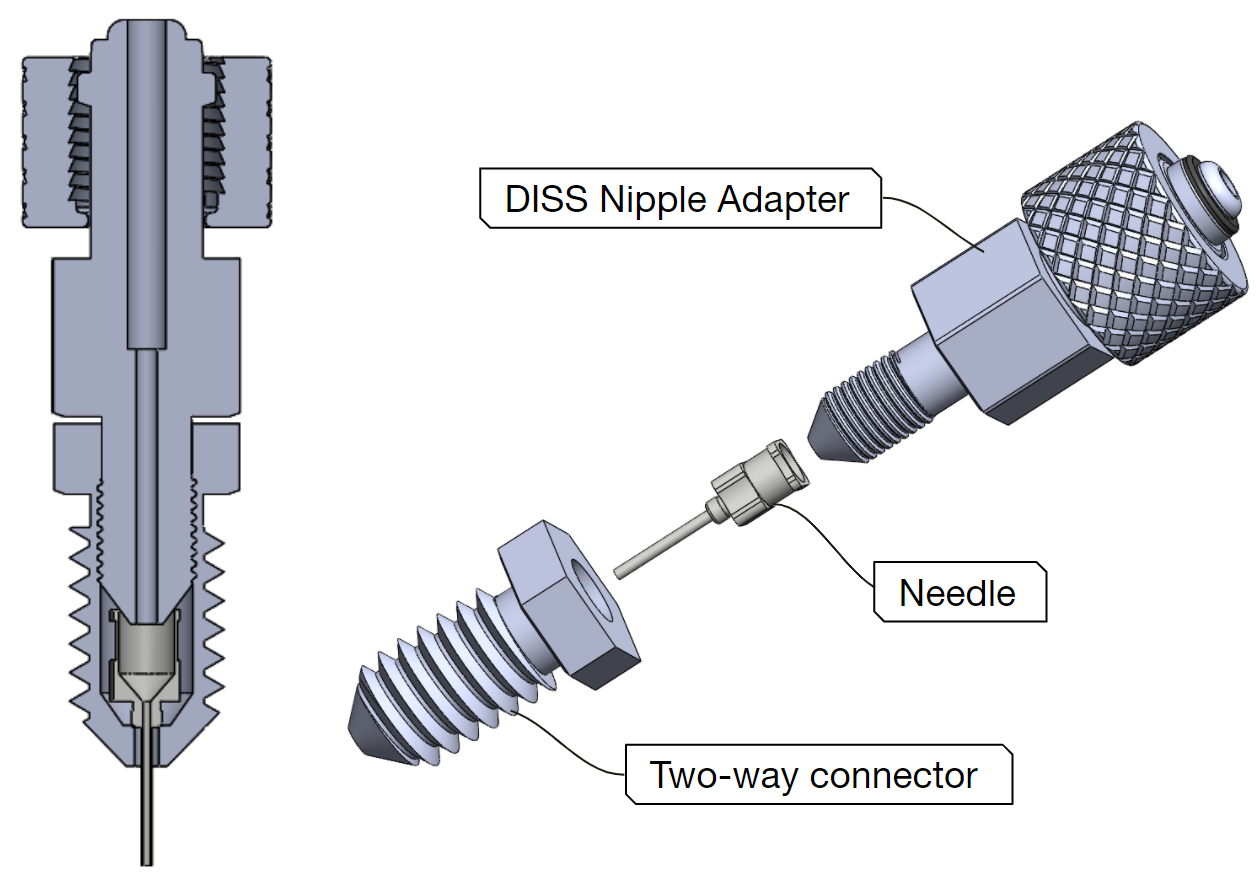}}
    \caption{Illustration of exploded and cross-sectional views of Assembled adapters with primary nozzle }
    \label{assembled-primary-nozzle}
\end{figure}

Fig \ref{assembled-primary-nozzle} illustrates the way a two-way metal connector attaches a needle (primary nozzle) air-tightly with an off the shelf DISS Nipple adapter.

\subsubsection{Oxyjet Body}
Oxyjet body is another important component of the flow generator. It utilizes the supersonic driving flow of $O_2$ generated by the primary nozzle to pull air from the environment and produce an oxygen rich homogeneous high flow. Oxyjet body has three chambers. The first chamber is the suction chamber, where the primary inlet, the air inlet and any additional secondary inlet opens up.The high velocity driving flow from the primary inlet creates a low-pressure region which entrains environmental air. The combined flow then passes through the second chamber (mixing chamber). It is a constant-area chamber where the flows mix together to produces a homogeneous high flow. The high flow then passes through the third chamber (diffuser), where the flow is decelerated to induce pressure recovery. The output flow is oxygen rich homogeneous high flow, and can provide pressure (PEEP).

We aimed to make the Oxyjet body simple and 3D printable so that it can be produced with minimum manufacturing complexity. We started the base design of the Oxyjet body as a typical constant-area gas ejector, with three chambers (suction, mixing and diffuser chamber), two inlets (primary inlet and air inlet) and one outlet. However, this is a complex 3D structure and hence required CFD simulations for optimization. In \cite{LIAO} LIAO, he performed several CFD simulations on single phase gas ejector to determine the optimum range of design parameters to achieve the optimal performance. The geometry and dimension for most cases were maintained within that optimum range. Initially, we optimized the design for the needle with gauge 19 (primary nozzle), having a needle diameter of $D_n $ = 0.686 mm .

\begin{figure}[!h]
    \centering
    \resizebox{.9\textwidth}{!}{\includegraphics{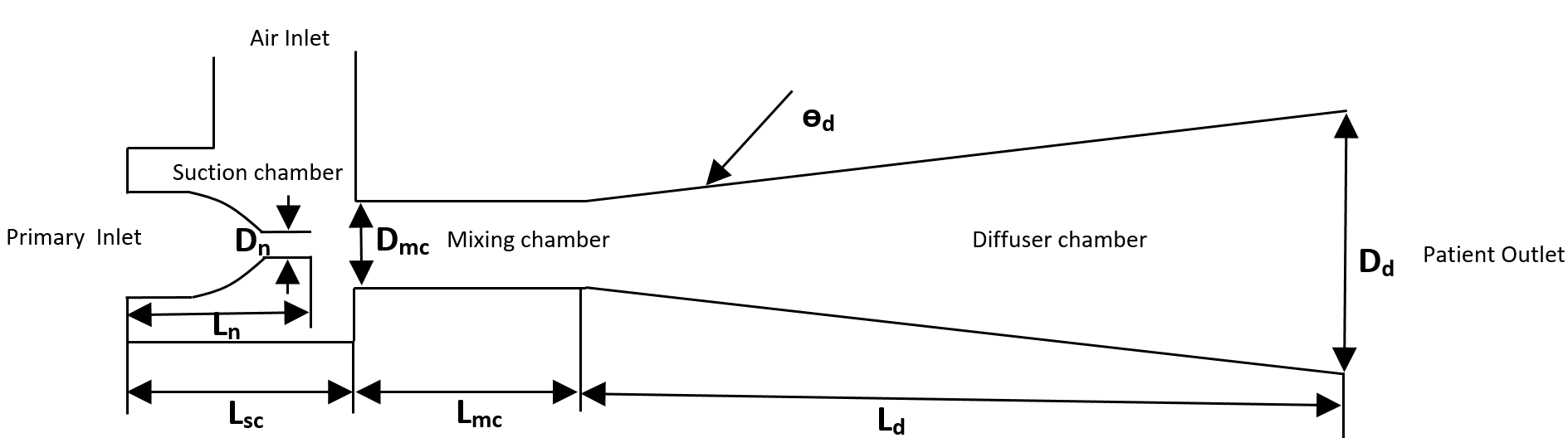}}
    \caption{Illustration of the diagram of a constant-area gas ejector}
    \label{ejector-diagram}
\end{figure}

In our design (as illustrated in Fig \ref{ejector-diagram}), we have used the diameter of the mixing chamber to be $D_{mc} = 6 mm $. The optimum range of diameter ratio of mixer to needle  $D_{mc}/D_n$ is between 8 to 14 \cite{LIAO}. And the diameter ratio of mixer to needle in our case is $D_{mc}/D_n = 8.75$, which is within the optimum range. There has to be a small gap between needle exit and mixing chamber, which is labeled as $L_{gap} = L_{sc} - L_n$. We have used $L_{gap} = 2.5 mm $. The optimum range of $L_{gap}/D_{mc}$ ratio is between 0.25 to 1.5 \cite{LIAO}. The ratio $L_{gap}/D_{mc}$ in our case is 0.42, which is also within the optimum range. For mixer length, we have used $L_{mc} = 30 mm$. In reality, the ratio of mixer length to mixer diameter $L_{mc}/D_{mc}$ is designed to be within the range of 8 to 12. However it was shown through CFD analysis, the air entrainment of the ejector decreases with the increase in  $L_{mc}/D_{mc}$ \cite{LIAO}. The ratio  $L_{mc}/D_{mc}$ in our case is 5, which is less than the suggested ratio. This choice was made to keep the device compact without hampering the performance. Finally, we have used the diffuser length to be $L_d = 60 mm $ and diffuser outlet diameter to be $D_d = 15 mm $. The optimum diffuser expansion angle is in the range of $2^{\circ}$ to $6^{\circ}$ \cite{LIAO}. The diffuser expansion angle $\theta_d = 4.3^{\circ}$ in our case, which is within the optimum range. The geometry details of the oxyjet body optimized for needle with gauge 19 is illustrated in Table \ref{table-geometry}.

\begin{table}[ht]
\centering
\begin{tabular}{l | l | l | l | l | l | l | l | l }
Needle Gauge & $D_n$(mm) & $L_n$(mm) & $L_{gap}$(mm) & $D_{mc}$(mm) & $L_{mc}$(mm) & $L_d$(mm) & $D_d$(mm) & $\theta_d$ \\
\hline
19 & 0.686 & 12.5 & 2.5 & 6 & 30 & 60 & 15 & 4.3 \\
\end{tabular}
\caption{Optimized geometry parameters of a gas ejector with needle gauge 19}
\label{table-geometry}
\end{table}

Maintaining the important geometry parameters within the optimum range we have modified other parts of the oxyjet body to make it a single piece 3D printable design. We have inclined the air inlet to an angle of 55$^\circ$ and set it's inner diameter to 16 mm and outer diameter to 22 mm, so that the pulled air can easily flow with the primary jet and we can easily attach a standard viral filter to the air inlet port. An additional inlet has been added at the suction chamber (first chamber) to supply more O$_2$ from a secondary O$_2$ source (via oxygen tube) and increase the \%FiO$_2$ (up to 100\%) when needed. This secondary inlet was also inclined to an angle of 50$^\circ$. We have designed the structure of the secondary inlet similar to the outlet of a christmas tree O$_2$ adapter. These inclinations are important for manufacturing the oxyjet body using 3D printing without support. There is threading on the inside wall at the primary inlet of the oxyjet body, which is used for the attachment of the primary nozzle arrangement (Fig \ref{assembled-primary-nozzle}). 

\begin{figure} [h]
        \centering
        \begin{tabular}{c c}
        \includegraphics[width=0.45\textwidth]{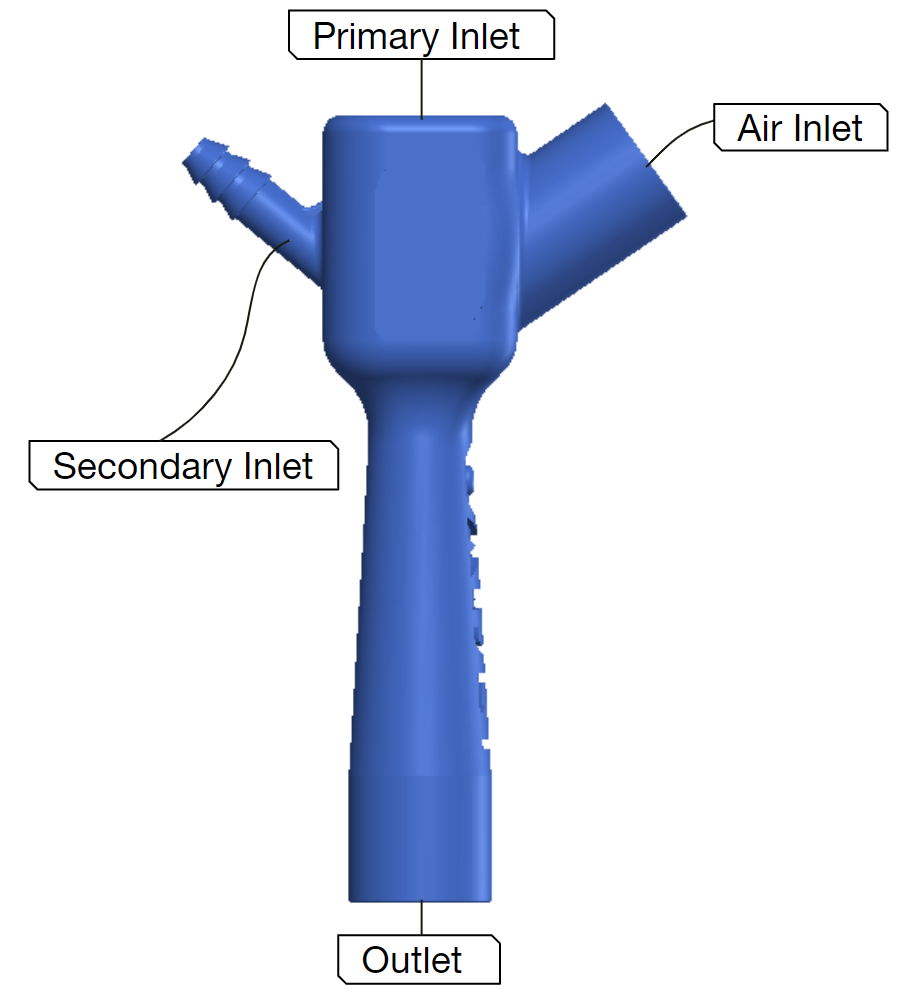}     &  \includegraphics[width=0.5\textwidth]{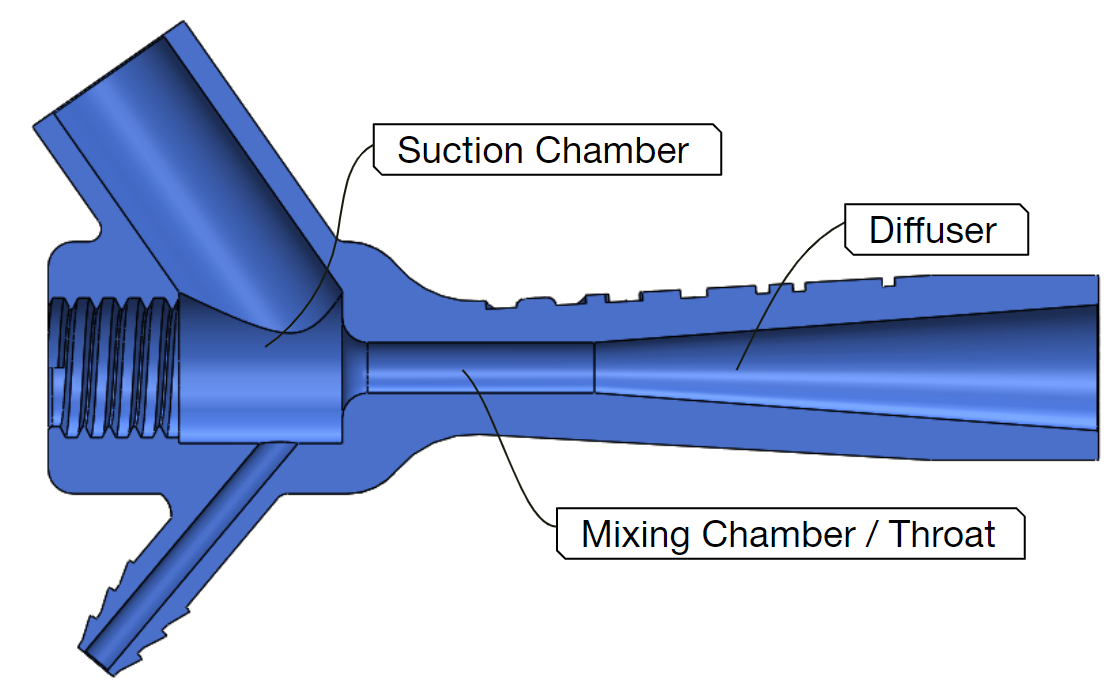}\\
         (a)    & (b)
        \end{tabular}
        \caption{Illustrations of (a) the oxyjet body and (b) its cross-sectional view.}\label{oxyjet_body}
\end{figure}

\subsubsection{Performance Evaluation of Oxyjet flow generator}
\begin{figure}[!h]
        \centering
        \includegraphics[width=0.75\textwidth]{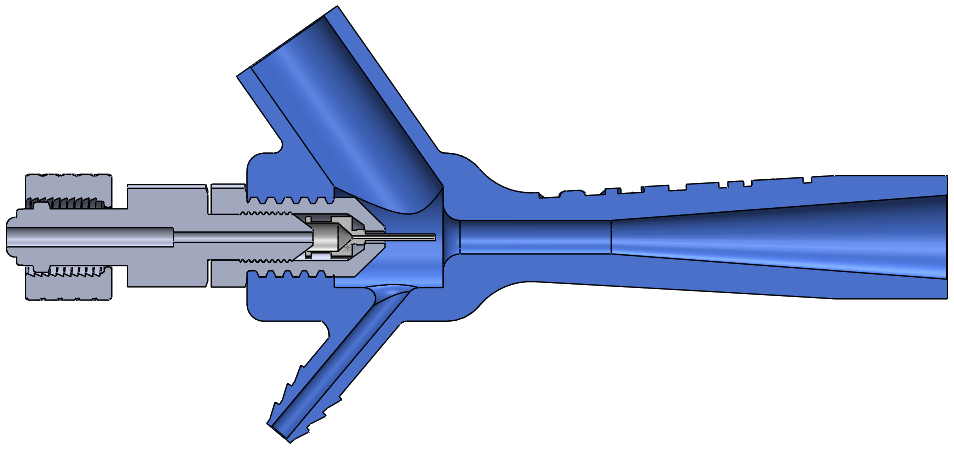}
        \caption{Illustration of the cross-sectional view of the oxyjet flow generator}
        \label{oxyjet_generator_cross_section}
\end{figure}

The complete oxyjet flow generator is a complex structure having multiple inlets. It is not possible to evaluate it's performance using one-dimensional flow analysis. Hence, we have to use CFD analysis and practical experiments for the device characterization. Fig \ref{oxyjet_generator_cross_section} illustrates a cross-sectional view of the oxyjet flow generator.

\begin{figure}[!h]
        \centering
        \resizebox{.7\textwidth}{!}{\includegraphics{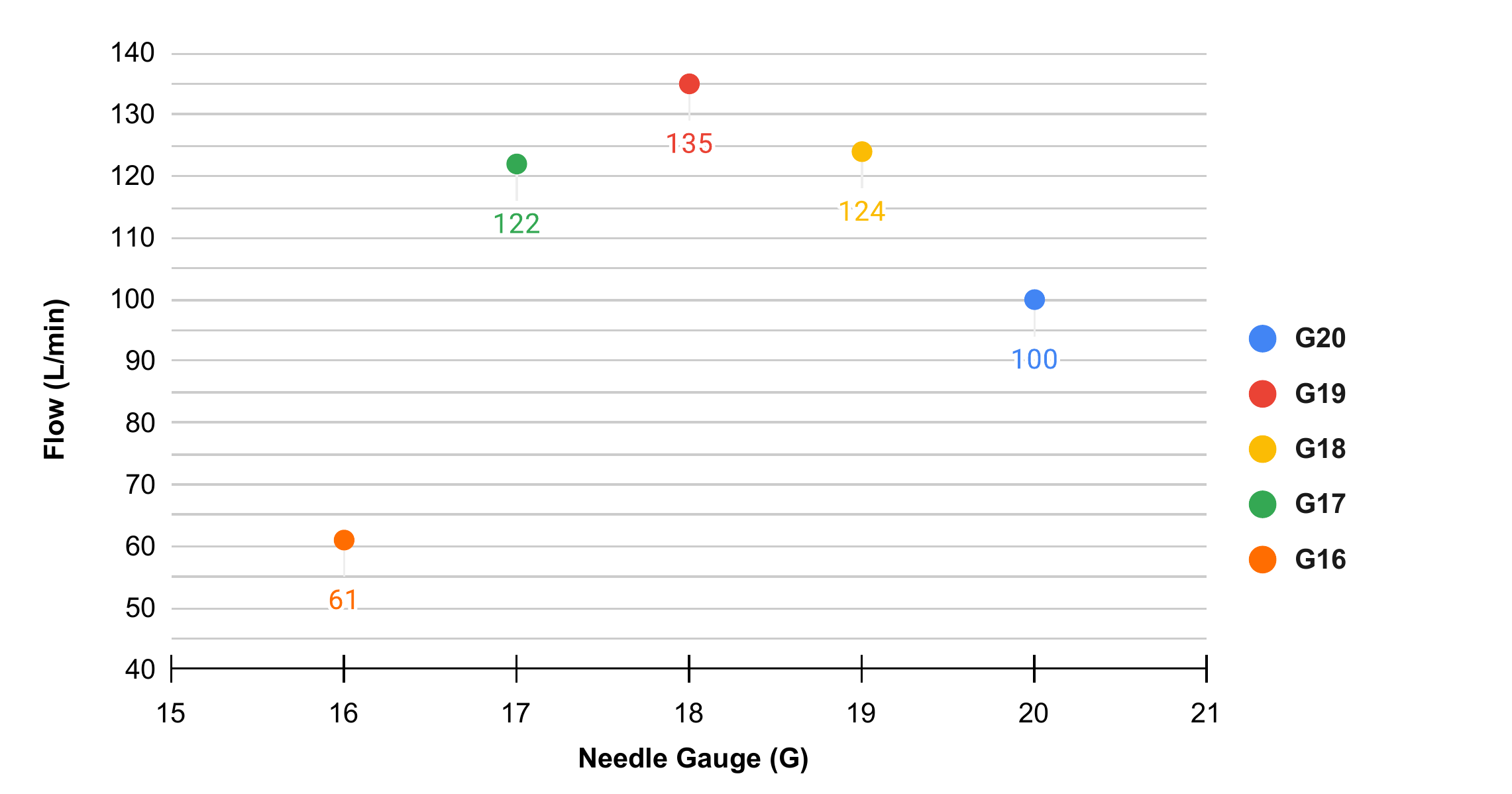}}
        \caption{Illustration of total output flow (L/min) of  oxyjet flow generator for different needle gauges at open circuit condition.}
        \label{max_needle_exp}
\end{figure}

We have performed experimental analysis with different needle gauges (G20, G19, G18, G17 and G16) to determine the output flow (L/min) of the flow generator at open circuit condition as illustrated in Fig \ref{max_needle_exp}. However, the primary $O_2$ flow (L/min) (i.e. driving flow) passing through the needles are not the same for all gauges. This is because of the upper limits (15L/min flow and 3.5 bar or 50 psi stagnation pressure) of a standard flowmeter. When attached to a standard flowmeter, the primary $O_2$ flow (L/min) passing through the needles with gauges G20, G19 and G18, attains supersonic state. However, for needles with gauges G17 and G16, the flow remains in the subsonic compressible region (based on analysis of Fig \ref{c_pressure_velocity_p}). The output flow (L/min) of the oxyjet flow generator is highest (135 L/min) for G18 needle. This is because, the choked flow (L/min) passing through G18 needle is maximum. For higher needle gauges (G19 and G20), the choked flow (L/min) decreases, resulting in a decrease in total output flow (L/min). For lower needle gauges (G17 and G16), the flow remains in the subsonic compressible region which also results in a drastic decrease in the total output flow (L/min).

\begin{figure}[!h]
        \centering
        \resizebox{.7\textwidth}{!}{\includegraphics{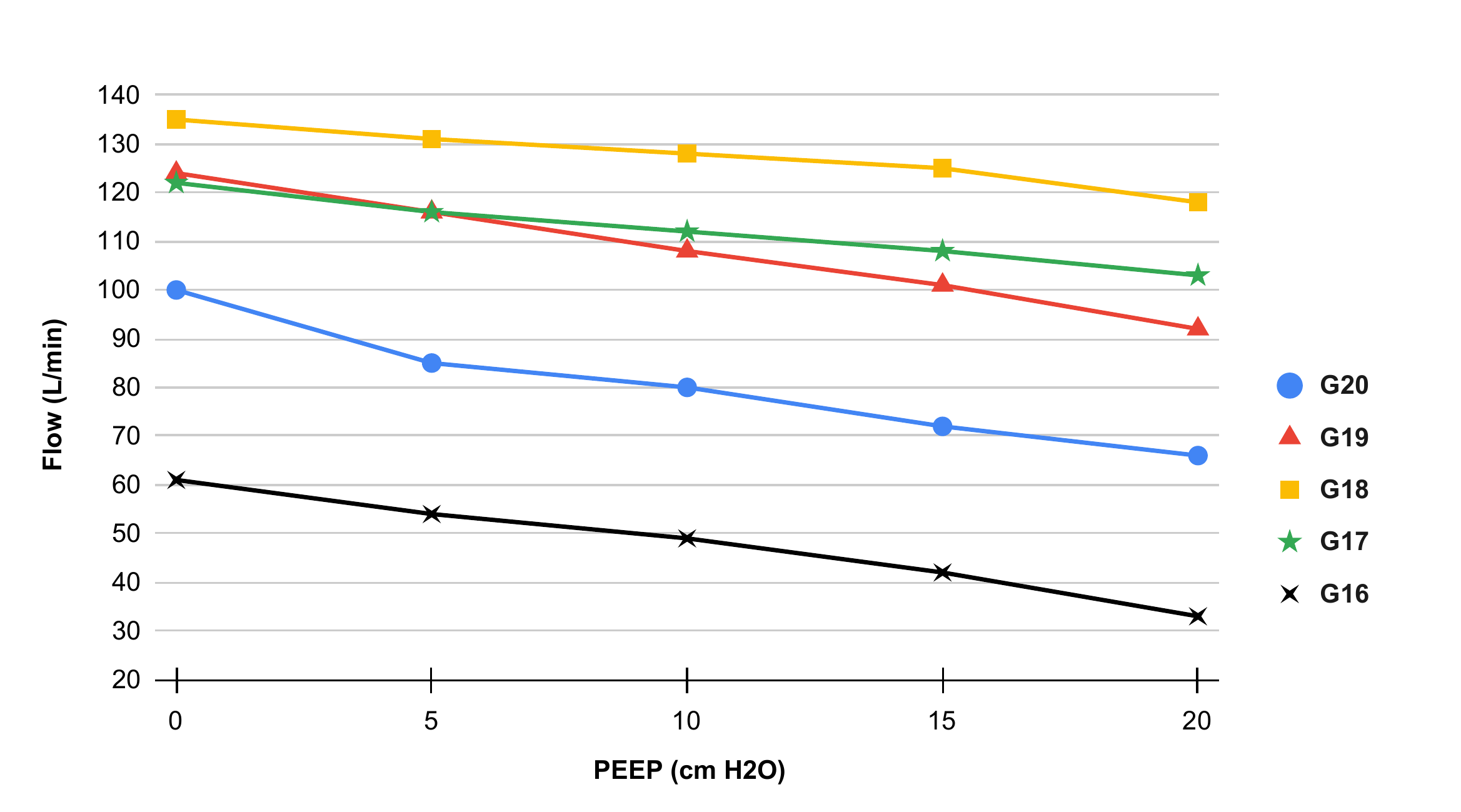}}
        \caption{Illustration of variation of total output flow (L/min) of the the oxyjet flow generator with PEEP (cm $H_2O$) for different needle gauges.}
        \label{totalflow_peep_exp}
\end{figure}

In the next step, we have evaluated the variation of output flow (L/min) of the flow generator with PEEP (cm $H_2O$) for different needle gauges as illustrated in Fig \ref{totalflow_peep_exp}. The output flow (L/min) decreases almost linearly with increasing PEEP for all needle gauges. The curve for G18 needle is above all, maintaining an output flow of 135 L/min and 118 L/min at 0 and 20 cm $H_2O$ PEEP respectively. The curve for G16 needle is below all, maintaining an output flow of 61L/min and 33 L/min at 0 and 20 cm $H_2O$ PEEP respectively. The curves of other needle gauges are in-between.

Based on experimental analysis we conclude: i) The flow generator can generate an output flow of 135 L/min with G18 needle at open circuit condition, which is it's highest flow ii) We can change the maximum output flow (L/min) of the flow generator by switching needle gauges.i) The flow generator can deliver pressurized high flow (upto 20 $cm H_2O$ PEEP) for all experimented needle gauges (G16, G17, G18, G19 and G20) iii) Flow generator with G16 needle delivers an output flow of 61 L/min at open circuit condition and 33 L/min at 20 cm $H_2O$ PEEP. Hence, flow generator with needle gauge lower than G16 should not be used in an early CPAP device, because the output  flow lower than this will be insufficient.

By now, we know the characteristic curves of the oxyjet flow generator and how its properties changes with needle gauges. Our next target is to minimize the oxygen consumption of the device. Minimizing oxygen consumption/waste is very important especially, for developing countries where there is a shortage of oxygen. Now, to do so, we need to lower the total output flow (L/min) of the flow generator so that we can increase $\%FiO_2$ of the total flow with less amount of $O_2$ flow (L/min). However, the total flow should not drop below the minimum requirements of early-stage COVID or hypoxemic patients. 

From the curves as illustrated in Fig \ref{totalflow_peep_exp} we can conclude, flow generator with G16 needle has the minimum oxygen consumption compared to other needle gauges and also fulfills minimum flow requirement criteria of early-stage COVID or hypoxemic patients. Hence, Flow generator with G16 needle is the viable choice for an early CPAP device in LMIC settings.

\begin{figure}[!h]
        \centering
        \resizebox{.7\textwidth}{!}{\includegraphics{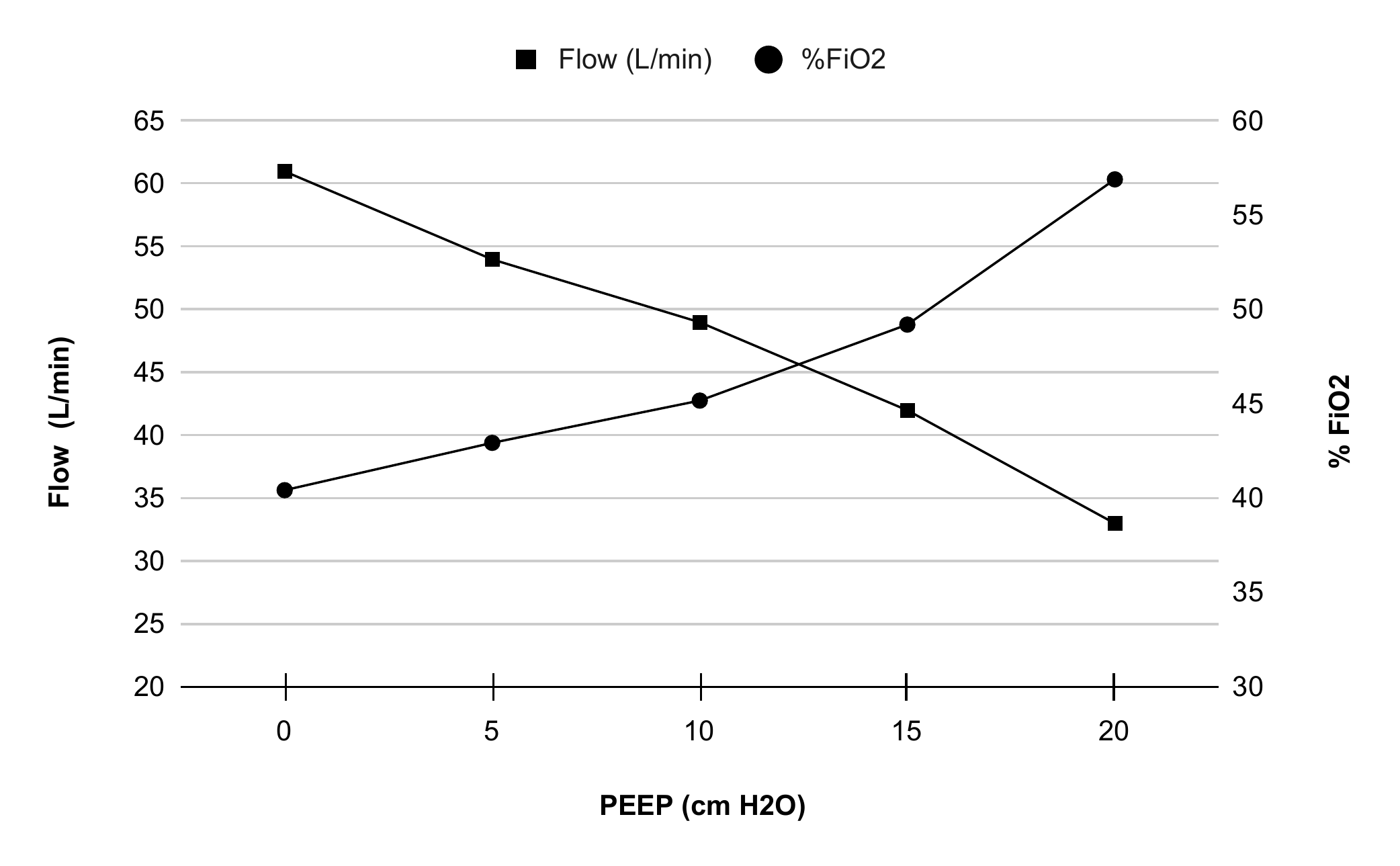}}
        \caption{Illustration of the maximum output flow (L/min) and minimum oxygen ($\%FiO_2$) delivered with PEEP (cm $H_2O$) by the oxyjet flow generator with G16 needle.}
        \label{G16_fio2_flow}
\end{figure}

We have performed further experimental analysis to evaluate the oxygen consumption of the flow generator with G16 needle. Fig \ref{G16_fio2_flow} illustrates the maximum output flow (L/min) and minimum oxygen ($\%FiO_2$) delivered by the oxyjet flow generator with G16 needle at different PEEP (cm $H_2O$). At open circuit condition, the maximum output flow and the minimum oxygen ($\%FiO_2$) of the flow generator are 61 L/min and 40\% respectively. The output flow decreases and the minimum oxygen ($\%FiO_2$) increases almost linearly with PEEP (cm $H_2O$). At 20 cm $H_2O$ PEEP, the output flow and the minimum oxygen ($\%FiO_2$) reach 33 L/min and 57\% respectively.

\begin{figure}[!h]
        \centering
        \resizebox{.7\textwidth}{!}{\includegraphics{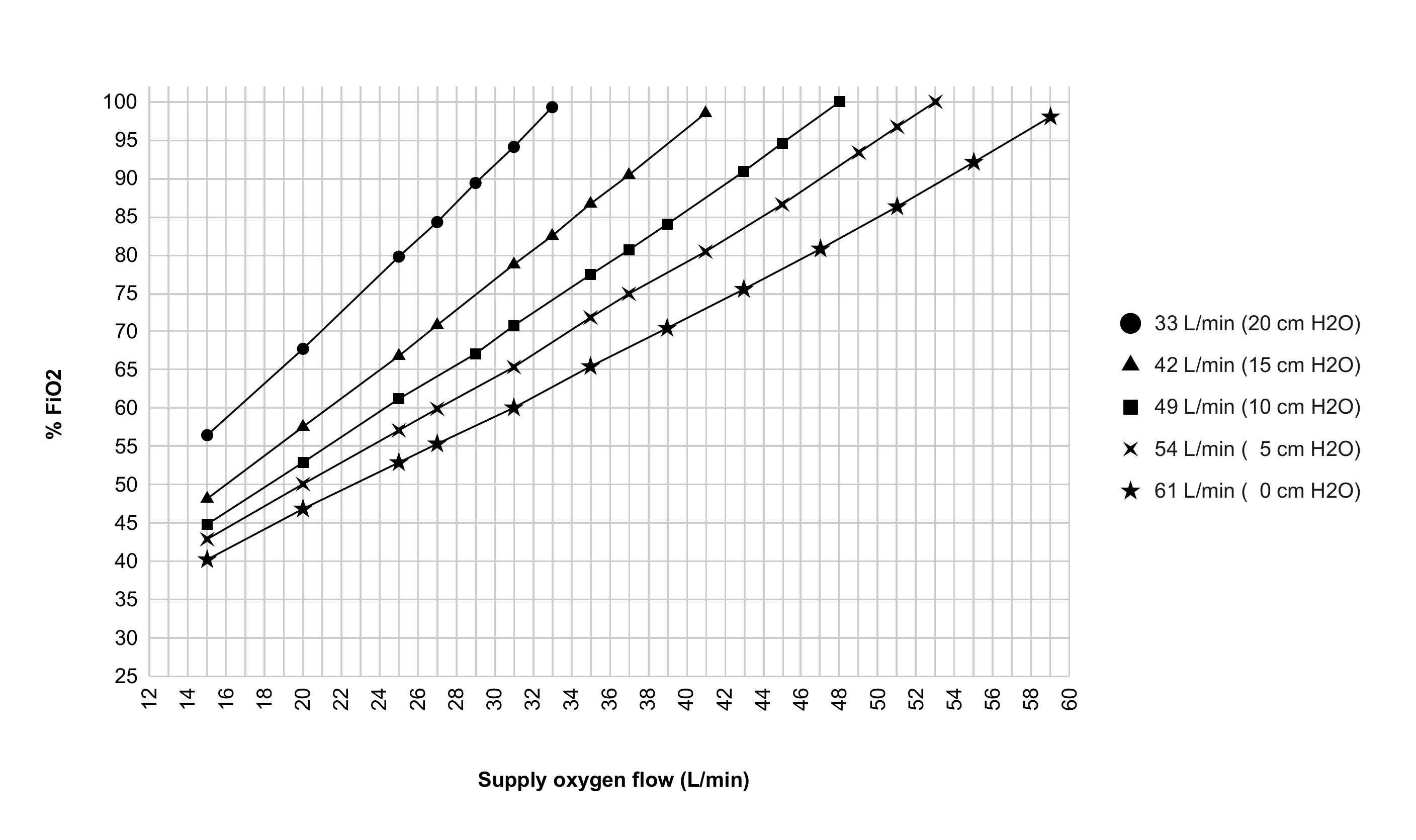}}
        \caption{Illustration of the oxygen consumption of the flow generator with G16 needle at different settings.}
        \label{G16_O2_consumption}
\end{figure}

Fig \ref{G16_O2_consumption} illustrates the oxygen consumption of the flow generator with G16 needle at different settings. For different PEEP (cm $H_2O$) setting, the $\%FiO_2$ follows a linear relation with the supplied oxygen flow (L/min) having different intercept and slope. At 0 cm $H_2O$ PEEP, the flow generator can provide 100\% oxygen concentration ($\%FiO_2$) using 61 L/min of supply oxygen via dual channel flowmeters (15L/min + 50 L/min). As the PEEP increases and the total output flow decreases, the amount of supply oxygen (L/min) to provide 100\% oxygen concentration ($\%FiO_2$) also decreases. 

\section{Conclusion}
This paper describes the design and preliminary evaluation of the Oxyjet CPAP system, particularly the flow-generator design. This is an ongoing project and this paper discloses an initial version of the design and evaluation of this device. Due to the pandemic and the need for low-cost oxygen delivery systems, we have made the decision to publish this pre-print and open our design for public use. However, we will continue to further develop and improve the device based on the ongoing clinical evaluation of the device.

\section{DISCLAIMER}
The OxyJet device design provided in this paper is an medical device prototype. It is not approved by any medical device regulatory authority and should not be used in the hospital without proper approval. Clinical evaluations of the device is currently ongoing using this device in Bangladesh with the approval of Bangladesh Medical Research Council (BMRC). The authors provide the design and evaluation data in this manuscript as is, and do not take responsibility for the failure of the device if manufactured by any one by using the designs provided herewith. If it is to be manufactured for medical use or clinical trials, it should be appropriately tested and evaluated according to the regulatory requirements that may apply.

\section{ACKNOWLEDGMENT}\label{sec:8}
We would firstly like to thank the Department of Biomedical Engineering (BME), BUET for supporting this research by providing laboratory access throughout the lockdown period. We would specifically thank Prof. Dr. Aynal Haque, Dr. Tarik Arafat and Dr. Jahid Ferdous. We are also grateful for the support of the BUET administration, specifically, Prof. Dr. Satya Prasad Majumder. We have had extensive discussions with medical doctors to understand the requirements of a CPAP device for Covid-19 patients. Among them, we would like to thank Dr. Abdullah Al Ahmed, Dr. Fuad Galib, Prof. Dr. Robed Amin, Dr. Forhad Uddin Hasan Choudhury, Dr. Khairul Islam and Dr. Mohiuddin Sharif.

\bibliography{main}

\begin{thebibliography}{10}
\urlstyle{rm}
\expandafter\ifx\csname url\endcsname\relax
  \def\url#1{\texttt{#1}}\fi
\expandafter\ifx\csname urlprefix\endcsname\relax\def\urlprefix{URL }\fi
\expandafter\ifx\csname doiprefix\endcsname\relax\def\doiprefix{DOI: }\fi
\providecommand{\bibinfo}[2]{#2}
\providecommand{\eprint}[2][]{\url{#2}}

\bibitem{world2020clinical}
\bibinfo{author}{{W H O}}.
\newblock \bibinfo{title}{{Clinical management of COVID-19: interim guidance,
  27 May 2020}}.
\newblock \bibinfo{type}{Tech. Rep.}, \bibinfo{institution}{{World Health
  Organization}} (\bibinfo{year}{2020}).

\bibitem{whittle2020respiratory}
\bibinfo{author}{Whittle, J.~S.}, \bibinfo{author}{Pavlov, I.},
  \bibinfo{author}{Sacchetti, A.~D.}, \bibinfo{author}{Atwood, C.} \&
  \bibinfo{author}{Rosenberg, M.~S.}
\newblock \bibinfo{journal}{\bibinfo{title}{{Respiratory support for adult
  patients with COVID-19}}}.
\newblock {\emph{\JournalTitle{Journal of the American College of Emergency
  Physicians Open}}} \textbf{\bibinfo{volume}{1}}, \bibinfo{pages}{95--101}
  (\bibinfo{year}{2020}).

\bibitem{rhodes2017surviving}
\bibinfo{author}{Rhodes, A.} \emph{et~al.}
\newblock \bibinfo{journal}{\bibinfo{title}{{Surviving sepsis campaign:
  international guidelines for management of sepsis and septic shock: 2016}}}.
\newblock {\emph{\JournalTitle{Intensive care medicine}}}
  \textbf{\bibinfo{volume}{43}}, \bibinfo{pages}{304--377}
  (\bibinfo{year}{2017}).

\bibitem{rimensberger2015ventilatory}
\bibinfo{author}{Rimensberger, P.~C.}, \bibinfo{author}{Cheifetz, I.~M.},
  \bibinfo{author}{Group, P. A. L. I. C.~C.} \emph{et~al.}
\newblock \bibinfo{journal}{\bibinfo{title}{Ventilatory support in children
  with pediatric acute respiratory distress syndrome: proceedings from the
  pediatric acute lung injury consensus conference}}.
\newblock {\emph{\JournalTitle{Pediatric Critical Care Medicine}}}
  \textbf{\bibinfo{volume}{16}}, \bibinfo{pages}{S51--S60}
  (\bibinfo{year}{2015}).

\bibitem{covid19guidelineBD}
\bibinfo{author}{{Directorate General of Health Services (DGHS)}}.
\newblock \bibinfo{title}{{National Guidelines on Clinical Management of
  Coronavirus Disease 2019 (COVID-19)}}.
\newblock \bibinfo{type}{Tech. Rep.}, \bibinfo{institution}{{Ministry of Health
  \& Family Welfare, Government of the People's Republic of Bangladesh}}
  (\bibinfo{year}{2020}).

\bibitem{oxygen2020news}
\bibinfo{author}{Mollah, S.} \& \bibinfo{author}{Saad, M.}
\newblock \bibinfo{title}{{Crisis of Oxygen Cylinders: Demand sky high, prices
  too}} (\bibinfo{year}{2020}).

\bibitem{abdullah2020number}
\bibinfo{author}{Abdullah, M.}
\newblock \bibinfo{title}{{Number of ICU beds insufficient to combat Covid-19
  pandemic}} (\bibinfo{year}{2020}).

\bibitem{mccarthy2020countries}
\bibinfo{author}{Forbes}.
\newblock \bibinfo{title}{The countries with the most critical care beds per
  capita [infographic]}.
\newblock
  \bibinfo{howpublished}{\url{https://www.forbes.com/sites/niallmccarthy/2020/03/12/the-countries-with-the-most-critical-care-beds-per-capita-infographic/\#317ad40c7f86}}.
\newblock \bibinfo{note}{(Accessed on 08/05/2020)}.

\bibitem{namendys2020respiratory}
\bibinfo{author}{{\~N}amendys-Silva, S.~A.}
\newblock \bibinfo{journal}{\bibinfo{title}{{Respiratory support for patients
  with COVID-19 infection}}}.
\newblock {\emph{\JournalTitle{The Lancet Respiratory Medicine}}}
  \textbf{\bibinfo{volume}{8}}, \bibinfo{pages}{e18} (\bibinfo{year}{2020}).

\bibitem{rochwerg2017official}
\bibinfo{author}{Rochwerg, B.} \emph{et~al.}
\newblock \bibinfo{journal}{\bibinfo{title}{{Official ERS/ATS clinical practice
  guidelines: noninvasive ventilation for acute respiratory failure}}}.
\newblock {\emph{\JournalTitle{European Respiratory Journal}}}
  \textbf{\bibinfo{volume}{50}} (\bibinfo{year}{2017}).

\bibitem{slessarev2020patient}
\bibinfo{author}{Slessarev, M.}, \bibinfo{author}{Cheng, J.},
  \bibinfo{author}{Ondrejicka, M.} \& \bibinfo{author}{Arntfield, R.}
\newblock \bibinfo{journal}{\bibinfo{title}{{Patient self-proning with
  high-flow nasal cannula improves oxygenation in COVID-19 pneumonia}}}.
\newblock {\emph{\JournalTitle{Canadian Journal of Anesthesia/Journal canadien
  d'anesth{\'e}sie}}} \bibinfo{pages}{1--3} (\bibinfo{year}{2020}).

\bibitem{lawton2020reduced}
\bibinfo{author}{Lawton, T.} \emph{et~al.}
\newblock \bibinfo{journal}{\bibinfo{title}{{Reduced ICU demand with early CPAP
  and proning in COVID-19 at Bradford: a single centre cohort}}}.
\newblock {\emph{\JournalTitle{medRxiv}}}  (\bibinfo{year}{2020}).

\bibitem{ashish2020early}
\bibinfo{author}{Ashish, A.} \emph{et~al.}
\newblock \bibinfo{journal}{\bibinfo{title}{{Early CPAP reduced mortality in
  covid-19 patients. Audit results from Wrightington, Wigan and Leigh Teaching
  Hospitals NHS Foundation Trust}}}.
\newblock {\emph{\JournalTitle{medRxiv}}}  (\bibinfo{year}{2020}).

\bibitem{duan2020use}
\bibinfo{author}{Duan, J.} \emph{et~al.}
\newblock \bibinfo{journal}{\bibinfo{title}{{Use of high-flow nasal cannula and
  noninvasive ventilation in patients with COVID-19: A multicenter
  observational study}}}.
\newblock {\emph{\JournalTitle{The American Journal of Emergency Medicine}}}
  (\bibinfo{year}{2020}).

\bibitem{nhs1559guidance}
\bibinfo{author}{NHS, B.}
\newblock \bibinfo{journal}{\bibinfo{title}{{Guidance for the role and use of
  non-invasive respiratory support in adult patients with coronavirus
  (confirmed or suspected) 2020 26 March 2020 Version 2}}}.
\newblock {\emph{\JournalTitle{Publications approval reference}}}
  \textbf{\bibinfo{volume}{1559}} (\bibinfo{year}{2020}).

\bibitem{And}
\bibinfo{author}{András~Lovas, D.~T., Márton Ferenc~Németh} \&
  \bibinfo{author}{Molnár, Z.}
\newblock \bibinfo{journal}{\bibinfo{title}{{Lung Recruitment Can Improve
  Oxygenation in Patients Ventilated in Continuous Positive Airway
  Pressure/Pressure Support Mode}}}.
\newblock {\emph{\JournalTitle{[Updated 2015 Apr 21]. Front Med (Lausanne)}}}
  (\bibinfo{year}{Publishing; 2015 Apr 21.}).

\bibitem{simonds2010evaluation}
\bibinfo{author}{Simonds, A.} \emph{et~al.}
\newblock \bibinfo{journal}{\bibinfo{title}{{Evaluation of droplet dispersion
  during non-invasive ventilation, oxygen therapy, nebuliser treatment and
  chest physiotherapy in clinical practice: implications for management of
  pandemic influenza and other airborne infections}}}.
\newblock {\emph{\JournalTitle{Health technology assessment (Winchester,
  England)}}} \textbf{\bibinfo{volume}{14}}, \bibinfo{pages}{131--172}
  (\bibinfo{year}{2010}).

\bibitem{ing2020role}
\bibinfo{author}{Ing, R.~J.} \emph{et~al.}
\newblock \bibinfo{journal}{\bibinfo{title}{{Role of Helmet-Delivered
  Noninvasive Pressure Support Ventilation in COVID-19 Patients}}}.
\newblock {\emph{\JournalTitle{Journal of Cardiothoracic and Vascular
  Anesthesia}}}  (\bibinfo{year}{2020}).

\bibitem{marini2020management}
\bibinfo{author}{Marini, J.~J.} \& \bibinfo{author}{Gattinoni, L.}
\newblock \bibinfo{journal}{\bibinfo{title}{{Management of COVID-19 respiratory
  distress}}}.
\newblock {\emph{\JournalTitle{Jama}}}  (\bibinfo{year}{2020}).

\bibitem{radovanovic2020helmet}
\bibinfo{author}{Radovanovic, D.} \emph{et~al.}
\newblock \bibinfo{journal}{\bibinfo{title}{{Helmet CPAP to treat acute
  hypoxemic respiratory failure in patients with COVID-19: a management
  strategy proposal}}}.
\newblock {\emph{\JournalTitle{Journal of clinical medicine}}}
  \textbf{\bibinfo{volume}{9}}, \bibinfo{pages}{1191} (\bibinfo{year}{2020}).

\bibitem{gattinoni2020covid}
\bibinfo{author}{Gattinoni, L.} \emph{et~al.}
\newblock \bibinfo{title}{{COVID-19 pneumonia: different respiratory treatments
  for different phenotypes?}} (\bibinfo{year}{2020}).

\bibitem{needle}
\bibinfo{title}{Stainless steel dispensing needles - specification}.
\newblock
  \bibinfo{howpublished}{\url{https://www.vapebureau.com.au/products/85mm-stainless-steel-dispensing-needles-8-10-12-gauge}}.

\bibitem{needlegauge}
\bibinfo{title}{Birmingham gauge}.
\newblock
  \bibinfo{howpublished}{\url{https://en.wikipedia.org/wiki/Birmingham_gauge}}.

\bibitem{fisher}
\bibinfo{author}{Solutions, E.~A.}
\newblock \bibinfo{title}{Material guidelines for gaseous oxygen service}.
\newblock
  \bibinfo{howpublished}{\url{https://www.emerson.com/documents/automation/product-bulletin-material-guidelines-for-gaseous-oxygen-service-en-141422.pdf}}
  (\bibinfo{year}{2017}).

\bibitem{LIAO}
\bibinfo{author}{LIAO, C.}
\newblock \bibinfo{journal}{\bibinfo{title}{Gas ejector modeling for design and
  analysis}}.
\newblock {\emph{\JournalTitle{Ph.D. Dissertation, Nuclear Engineering Dept.,
  Texas A\&M Univ., College Station, TX}}}  (\bibinfo{year}{2008}).

\end{thebibliography}

\end{document}